\documentclass[lettersize,journal]{IEEEtran}
\usepackage{amsmath,amsfonts}
\usepackage{algorithmic}
\usepackage{array}
\usepackage[caption=false,font=normalsize,labelfont=sf,textfont=sf]{subfig}
\usepackage{textcomp}
\usepackage{stfloats}
\usepackage{url}
\usepackage{verbatim}
\usepackage{graphicx}
\def\BibTeX{{\rm B\kern-.05em{\sc i\kern-.025em b}\kern-.08em
    T\kern-.1667em\lower.7ex\hbox{E}\kern-.125emX}}
\usepackage{balance}

\usepackage{orcidlink}
\usepackage{kotex}
\usepackage{mathptmx}
\usepackage{booktabs}
\usepackage{multirow}
\usepackage{amsmath}
\usepackage{xcolor}

\graphicspath{{figs/}{figures/}{pictures/}{images/}{./}} 

\definecolor{lightyellow}{RGB}{255,255,180}
\definecolor{lightorange}{RGB}{255,230,180}

\begin{document}

\title{OPRA-Vis: Visual Analytics System to Assist Organization-Public Relationship Assessment with Large Language Models}
\author{
    Sangbong Yoo\orcidlink{0000-0002-0973-9288},~\IEEEmembership{Member,~IEEE}, Seongbum Seo\orcidlink{0000-0002-9582-1674}, Chanyoung Yoon\orcidlink{0000-0002-9784-0238}, Hyelim Lee\orcidlink{0000-0001-9032-9835}, Jeong-Nam Kim\orcidlink{0000-0002-4254-5784}, Chansoo Kim, Yun Jang\orcidlink{0000-0001-7745-1158},~\IEEEmembership{Member,~IEEE}, and Takanori Fujiwara\orcidlink{0000-0002-6382-2752},~\IEEEmembership{Member,~IEEE}
\thanks{S. Yoo and C. Kim are with AI, Information and Reasoning (AI/R) Laboratory, Korea Institute of Science and Technology (KIST), Seoul, South Korea. e-mail: \{usangbong; eau\}@kist.re.kr}
\thanks{S. Seo, C. Yoon, and Y. Jang are with Sejong University, Seoul, South Korea. E-mail: seo@seongbum.com, vfgtr8746@gmail.com, jangy@sejong.edu.}
\thanks{H. Lee is with Korea University, Seoul, South Korea. E-mail: hyelim\_lee@korea.ac.kr.}
\thanks{J.-N. Kim is with Korea Advanced Institute of Science and Technology (KAIST), Daejeon, South Korea. E-mail: layinformatics@gmail.com.}
\thanks{T. Fujiwara is with the University of Arizona. E-mail: tfujiwara@arizona.edu.}
\thanks{S. Yoo and S. Seo contributed equally to this work.}
\thanks{Y. Jang is the corresponding author.}
}

\markboth{IEEE Transactions on Visualization and Computer Graphics,~Vol.~xx, No.~xx, xx~20xx}%
{Yoo \MakeLowercase{\textit{et al.}}: OPRA-Vis: Visual Analytics System to Assist Organization-Public Relationship Assessment with Large Language Models}

\maketitle

\begin{abstract}
Analysis of public opinions collected from digital media helps organizations maintain positive relationships with the public. Such public relations (PR) analysis often involves assessing opinions, for example, measuring how strongly people trust an organization. Pre-trained Large Language Models (LLMs) hold great promise for supporting Organization-Public Relationship Assessment (OPRA) because they can map unstructured public text to OPRA dimensions and articulate rationales through prompting. However, adapting LLMs for PR analysis typically requires fine-tuning on large labeled datasets, which is both labor-intensive and knowledge-intensive, making it difficult for PR researchers to apply these models.
In this paper, we present OPRA-Vis, a visual analytics system that leverages LLMs for OPRA without requiring extensive labeled data. Our framework employs Chain-of-Thought prompting to guide LLMs in analyzing public opinion data by incorporating PR expertise directly into the reasoning process. Furthermore, OPRA-Vis provides visualizations that reveal the clues and reasoning paths used by LLMs, enabling users to explore, critique, and refine model decisions. We demonstrate the effectiveness of OPRA-Vis through two real-world use cases and evaluate it quantitatively, through comparisons with alternative LLMs and prompting strategies, and qualitatively, through assessments of usability, effectiveness, and expert feedback.
\end{abstract}


\section{Introduction}
\label{sec:intro}
\IEEEPARstart{C}{itizens} actively express their thoughts and emotions about issues they face~\cite{kim2011problem}. Social organizations, such as corporations and governments, influence individuals, while these individuals, including publics and stakeholders, also shape organizational actions~\cite{kim2013strategic, grunig2017publics}. Their communicative behaviors generate opinion data~\cite{kim2014lay, kim2021pseudo}, reflecting experiences and revealing shifts in relationship quality. Understanding such data provides insights into public expectations, aiding organizations in strategy, policy, and legitimacy management~\cite{grunig2002excellent}.

Public relations (PR), a discipline of strategic management, relies on analyzing public expectations to maintain legitimacy and secure resources. Traditionally, researchers have used Organization-Public Relationship Assessment~(OPRA)~\cite{hon1999guidelines, huang2001opra} with Likert scale surveys to quantify the public's relationship with organizations~\cite{kim2014lay, kim2013integrating}. However, these self-reports are prone to recall bias and social desirability, and cannot fully capture the dynamic nature of relationships due to their one-time format. To overcome these limitations, scholars are increasingly turning to social media and other online platforms, producing abundant textual data reflecting public opinion~\cite{kim2014lay, kim2021pseudo}. Yet, interpreting such online texts poses linguistic challenges, such as jargon, quotations, and missing context, and requires converting unstructured input into OPRA's structured dimensions. These difficulties require adopting Natural Language Processing (NLP) techniques to analyze text on a scale while preserving theoretical alignment with PR constructs.

Large Language Models (LLMs) support diverse NLP tasks and are widely adopted through pre-trained language models (PLMs)~\cite{beltagy2019scibert, gu2022domain}. While PLMs reduce the burden of training from scratch, integrating PR theory into LLMs for public opinion analysis faces three challenges. \textbf{C1:} LLMs require large labeled datasets for optimal performance on new tasks~\cite{beltagy2019scibert, howard2018universal}. \textbf{C2:} Labeling based on PR theory demands labor-intensive human coding~\cite{kim2011problem, grunig2001guidelines}. \textbf{C3:} LLMs provide limited transparency in how they map unstructured text to OPRA dimensions, making it challenging to confirm or adjust the rationale behind their relational judgments~\cite{wang2021want, sun2023text, zhao2024explainability}.

To tackle the three challenges (C1--C3), we introduce OPRA-Vis, a novel visual analytics (VA) system that integrates OPRA concepts of PR, LLMs, and visual analysis. The assessment and reasoning of PR experts is replicated by the system through the use of Gemma~\cite{team2024gemma} and Chain-of-Thought (CoT) prompting with clues and reasons~\cite{sun2023text}, based on \textit{few-shot examples (FSEs)}~\cite{brown2020language}. Bidirectional Encoder Representations from Transformers (BERT)~\cite{devlin2018bert} is used to classify the sentiment of words in sentences, thereby providing PR analysis with insights into the linguistic phenomena that LLMs comprehend. Visual analysis modules interpret and interact with LLM decision processes through scatter plots with tag clouds, attention visualization, and prompt-editing interactions. We evaluate OPRA-Vis using two real-world datasets, conduct quantitative comparisons with baseline classifiers, and provide qualitative assessments through expert feedback and usability tests. Our main contributions are summarized as follows.
\begin{itemize}
    \item We propose OPRA-Vis, an interactive visualization integrating PR theory (e.g., OPRA) and LLMs for labeling.
    \item We present an OPRA-informed continuous encoding and use a gravity model to visualize multiple concepts.
    \item We allow users to adjust LLM-generated labels by exposing the clues and reasoning behind the labeling procedure.
\end{itemize}
All supplementary materials, including the demonstration video, source codes, and LLM prompts used in OPRA-Vis, are available at \url{https://github.com/DVL-Sejong/OPRA-Vis}.
\section{Related Work}
In this section, we review the underlying techniques, including LLMs, VA for reasoning, and PR.

\subsection{Large Language Models (LLMs)}
Various state-of-the-art (SOTA) pre-trained LLMs, such as BERT~\cite{devlin2018bert} and Generative Pre-trained Transformers (GPT)~\cite{achiam2023gpt}, have demonstrated effectiveness across a wide range of NLP tasks. These models are typically fine-tuned for task-specific applications while retaining their original architectures. For example, SCIBERT~\cite{beltagy2019scibert} adapts BERT to scientific domains. To reduce fine-tuning costs, ULMFiT~\cite{howard2018universal} introduced transfer learning for NLP, though its generality is limited. Prompting methods, such as in-context learning~\cite{dong2022survey, chen2022improving}, guide LLMs via task \textit{descriptions} and \textit{examples}, enabling few- or zero-shot learning. CoT prompting~\cite{wei2022chain, wu2023chain} improves by incorporating reasoning steps into the prompt, but remains highly sensitive to prompt design~\cite{perez2021true}.

LLMs increasingly automate text classification~\cite{gu2022domain, ammar2018construction}, aiming to reduce human labeling efforts. Meng et al.~\cite{meng2020text} proposed self-training without labeled data, while Wang et al.~\cite{wang2021want} combined GPT-generated pseudo labels with human annotations. However, in domain-specific PR analysis, in-context learning still faces reliability challenges. Our approach differs by integrating expert-guided CoT prompting with interactive visualizations that expose model rationales and support prompt and label refinement for OPRA.

\subsection{Visual Analytics for Reasoning from Text Data}
VA facilitates effective decision-making by offering insights into data interpretation and human knowledge transformation~\cite{keim2008visual}. Tag clouds are a widely used method for identifying topics based on word frequency~\cite{liu2019bridging}. Knittel et al.~\cite{knittel2021pyramidtags} introduced an interactive tag cloud visualization that captures both temporal and semantic dynamics. Such approaches have been applied to text labeling analysis~\cite{khayat2020vassl, li2024evovis}. For example, Khayat et al.~\cite{khayat2020vassl} developed VASSL, a system for labeling social spambots---automated entities that influence human behavior---by combining dimensionality reduction, sentiment analysis, and topic modeling. VASSL incorporates tag clouds for topic analysis and provides controls over dimensionality reduction parameters to detect malicious behavior. Similarly, Li et al.~\cite{li2024evovis} proposed EvoVis for multi-class labeling, enabling users to explore relationships over historical overviews and supporting iterative analysis of labeling tasks.

Interactive VA systems also facilitate the interpretation of model decisions~\cite{vig2019multiscale, yeh2024attentionviz, coscia2023knowledgevis}. BertViz~\cite{vig2019multiscale} offers multiscale visualizations of transformer attention heads, neurons, and model structures, although it does not support interactive editing. AttentionViz~\cite{yeh2024attentionviz} visualizes joint query-key embeddings to reveal attention trends, but requires substantial pre-computation for large models. KnowledgeVIS~\cite{coscia2023knowledgevis} compares fill-in-the-blank prompts by generating semantic variations and interactive views, yet it has limitations in capturing fine-grained linguistic nuances. PrompTHis~\cite{guo2024prompthis} investigates how editing prompts influences image generation by highlighting word-level changes that affect the outputs.

These studies use LLMs and prompts to examine how input prompts affect model behavior. In contrast, our approach leverages prompting to overcome the scarcity of labeled data. Specifically, we utilize \textit{expert instructions} combined with CoT prompting that includes clues and reasons to embed domain knowledge into LLMs. Additionally, we visualize how these clues influence the model's reasoning for labeling.

\subsection{Public Relations (PR)}
The industry increasingly adopts digital media monitoring to track public sentiments and opinions toward organizations. Tools like Brandwatch~\cite{brandwatch} and other social media monitoring platforms support this process~\cite{hayes2021can, mohamed2022analyzing}. Academic researchers also use social network analysis to examine the evolution of public opinion. For instance, Himelboim et al.~\cite{himelboim2014asocial} analyzed Twitter followership to identify social mediators linking the U.S. Department of State with international publics, revealing that informal actors primarily mediated relationships in the Middle East and North Africa.

PR researchers also study public opinion on politics and government~\cite{kim2011problem, seltzer2010toward, lee2013explicating, chon2021predicting}. Political communication analysis in PR involves supporting initiatives of political parties, governments, and civil organizations through communication strategies~\cite{seltzer2010toward}. It also helps mitigate public concerns on social issues~\cite{chon2021predicting}. Chon and Park~\cite{chon2021predicting} examined OPR during disease outbreaks, aligning behavioral intentions with the situational theory of problem solving~\cite{kim2011problem} to promote Centers for Disease Control and Prevention (CDC) guideline adherence. PR studies further suggest that governments have a greater experience in crisis handling than citizens~\cite{lee2013explicating}. Lee and Jun~\cite{lee2013explicating} provided a theoretical view of public diplomacy, explaining how the U.S. and Korea foster cooperative relationships through international public experiences. This analysis also extends to public opinions from areas such as volunteerism~\cite{bortree2010exploring}.

Despite growing interest in artificial intelligence (AI), academic research applying deep learning or LLMs in PR remains scarce. Existing studies rely primarily on content analysis~\cite{mccorkindale2010can, linvill2012colleages} or case studies~\cite{andohquainoo2015theuseof}. A key obstacle is the lack of labeled data required to train such large AI models. In PR, expert-driven human coding remains essential. To address this challenge, we adopt LLMs and CoT prompting with clues and reasons to emulate expert reasoning by using structured \textit{FSEs} for assessing public opinion data.
\section{Design Considerations}
\label{sec:design}
This section presents the definitions of OPRA concepts, expert requirements, and the design goals of OPRA-Vis.

\subsection{Organization-Public Relationship Assessment (OPRA)}
\label{subsec:opra}
The OPRA~\cite{hon1999guidelines} is a framework devised by scholars in PR to evaluate relationships, featuring four fundamental concepts~\cite{grunig2001guidelines}: \textit{Trust}, \textit{Satisfaction}, \textit{Commitment}, and \textit{Control Mutuality}. \textit{Trust} comprises competence, dependability, and integrity. \textit{Satisfaction} relates to positive emotional responses. \textit{Commitment} concerns long-term expectations, while \textit{Control Mutuality} addresses perceptions of balanced control. OPRA provides metrics for gauging these relationships, offering valuable insight to maintain favorable public connections. These metrics are typically obtained via surveys, supporting relationship improvement strategies across sectors such as business and government, and helping organizations strengthen public engagement over both short- and long-term horizons. In PR analysis, OPRA concepts are evaluated at the sentence level. A sentence is labeled \textit{True} for a concept if it contains identifiable clues supporting that concept; otherwise, it is labeled \textit{False}. These binary labels denote the presence or absence of a concept, rather than the overall quality of the relationship. PR experts rely on cited clues and the distribution of \textit{True}/\textit{False} sentences to inform policy recommendations for strengthening organization–public relationships. Sentiment analysis complements OPRA labeling by capturing the emotional tone of a sentence (e.g., \textit{Positive} or \textit{Negative}), which is not conveyed by concept labels.

\subsection{Expert Requirements}
\label{subsec:ER}
Labeling public opinion data with OPRA concepts requires human coding by PR experts, which is resource-intensive. To mitigate this burden, we conducted semi-structured meetings with two PR experts. We synthesized a non-exhaustive but sufficient set of requirements to guide our design, with both experts later joining the project as coauthors.
\begin{itemize}
    \item [\textbf{R1}] PR experts lack labeled data and therefore need techniques that do not rely on training datasets.
    \item [\textbf{R2}] They need a labeling tool that uses PR expertise to replace the inconsistent and labor-intensive human coding.
    \item [\textbf{R3}] They aim to utilize OPRA concepts and sentiment for a detailed examination of ambiguous cases.
    \item [\textbf{R4}] They require information that allows them to assess whether the model's reasoning for labeling is reasonable.
    \item [\textbf{R5}] They require simple, intuitive interactions to correct labeling errors and refine the reasoning for labeling.
    \item [\textbf{R6}] They need a system that incorporates their edits across the remaining OPRA tasks to save time and effort.
\end{itemize}

\subsection{Design Goals}
The system design goals for the conjunction use of PR and LLM are as follows.
\begin{itemize}
    \item [\textbf{G1}] LLMs leverage PR theory using a small number of \textit{exemplars} (C1 \& C2 in Section~\ref{sec:intro}, R1, R2 in Section~\ref{subsec:ER}).
    \item [\textbf{G2}] To contextualize ambiguous cases, the system quantifies the certainty of OPRA-related clues in the text and classifies the sentiment of sentences (C2, C3, R3).
    \item [\textbf{G3}] The system presents the LLM's reasoning and decision process (C3, R4).
    \item [\textbf{G4}] Interactions are needed for PR experts to correct LLM decisions and errors (C2, C3, R5).
    \item [\textbf{G5}] The system enhances usability through task-relevant filtering informed by OPRA labeling (C2, C3, R6).
\end{itemize}

Integrating PR theory with LLMs for labeling public opinion data faces main challenges (C1–C3) as introduced in Section~\ref{sec:intro}. To address the challenge of insufficient training data for fine-tuning LLMs (G1), we implement the Gemma~\cite{team2024gemma} with CoT prompting, based on in-context learning. This method leverages \textit{FSEs} provided by PR experts to label the data. Moreover, we have developed an LLM pipeline that represents OPRA-related clues in public opinion data as a continuous encoding (G2). Adopting a BERT~\cite{devlin2018bert} reduces the dependence on human effort to classify sentiment by analyzing public opinion sentences. Sentiment classification enriches labeled data and broadens the analyses supported by our system (G2).

We combine VA with CoT prompting that incorporates clues and reasons to audit and revise the model's reasoning for labeling, supported by coordinated views and prompt editing (G3, G4, G5). CoT prompting elicits ``think-aloud'' rationales, making the LLM's reasoning for labeling explicit and auditable (G3). Additionally, understanding the transformer's attention mechanism is crucial for comprehending the LLM's reasoning for labeling and for guiding prompt editing in CoT-based revisions. Therefore, attention visualization is devised to provide an attention mechanism for the transformer that influences LLM's labeling (G3, G4). Moreover, public opinion data, i.e., freeform text posted on digital media, is diverse and noisy. Therefore, techniques that support usability are required for labeling, reasoning, and editing. We adopt sentence embeddings and dimensionality reduction to encode the certainty relevant to the OPRA concept (G2). This encoded certainty is then visualized in a scatter plot, with filtering applied based on certainty levels. The continuous encoding with certainty-based filtering for OPRA labeling improves usability and reduces human workload in visual analysis (G5).
\begin{figure}[t]
    \centering
    \includegraphics[width=\linewidth]{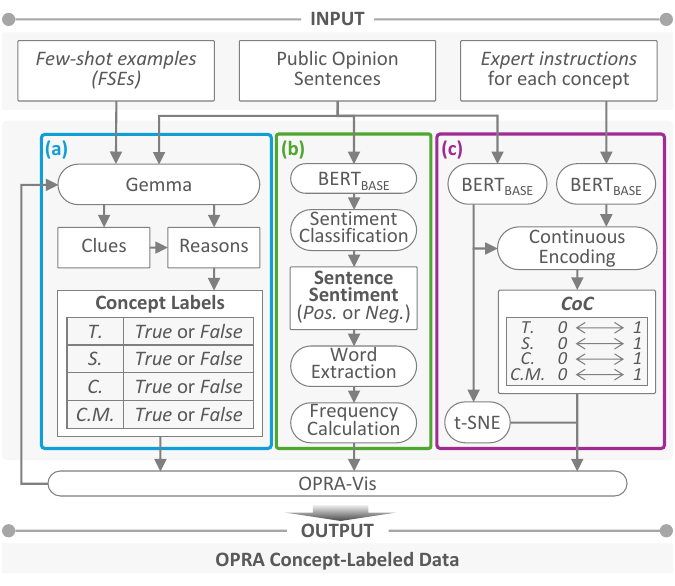}
    \caption{%
        Dataflow of OPRA-Vis. OPRA-Vis takes \textit{FSEs}, \textit{expert instructions} for each concept, and public opinion sentences as input and produces OPRA concept-labeled data. (a) shows initial concept labeling (\textit{True} or \textit{False}), (b) shows sentiment classification of sentences followed by frequency calculation, and (c) computes a certainty of concepts ($CoC$). \textit{T.}, \textit{S.}, \textit{C.}, and \textit{C.M.} are \textit{Trust}, \textit{Satisfaction}, \textit{Commitment}, and \textit{Control Mutuality}, respectively.
    }
    \label{fig:flow_single}
\end{figure}

\section{Data Processing of OPRA-Vis}
\label{sec:backend}
OPRA-Vis operates on an expert-in-the-loop workflow. The system first obtains initial LLM-generated concept labels (Section~\ref{subsec:opra}) and uses the certainty of concepts ($CoC$) to select review subsets by filtering. Experts update the prompt by editing \textit{few-shot examples} (\textit{FSEs}), after which the system re-evaluates the selected review subsets to revise labels and rationales. Iterating this workflow reduces the per-sentence review effort and progressively aligns decisions with expert criteria. For this semi-automatic workflow, we employ two LLMs---Gemma~(7B) and BERT$_\text{BASE}$~(110M)---as shown in Figure~\ref{fig:flow_single}: (a) concept labeling with clues and reasons, (b) sentiment classification followed by topic modeling, and (c) $CoC$ calculation. This section details the data-processing pipeline.

\subsection{Concept Labeling with Clues and Reasons}
\label{subsec:polarity_sentiment}
We begin by assigning OPRA concept labels to the unlabeled public-opinion sentences, as illustrated in Figure~\ref{fig:flow_single}~(a). Each sentence is designated as \textit{True} or \textit{False} for every OPRA concept using Gemma~(7B). Following the expert labeling, \textit{True} indicates the presence of identifiable clues supporting that concept within the sentence, while \textit{False} indicates their absence. To facilitate this labeling, we utilize CoT with clues and reasons (CoT$_\text{CR}$)~\cite{sun2023text}: \textit{FSEs} feature concept labels provided by experts, \texttt{Clues} serve as evidence relevant to the concepts, and \texttt{Reasons} offer concise rationales based on those clues.

\begin{figure*}[t]
    \centering
    \includegraphics[width=\linewidth]{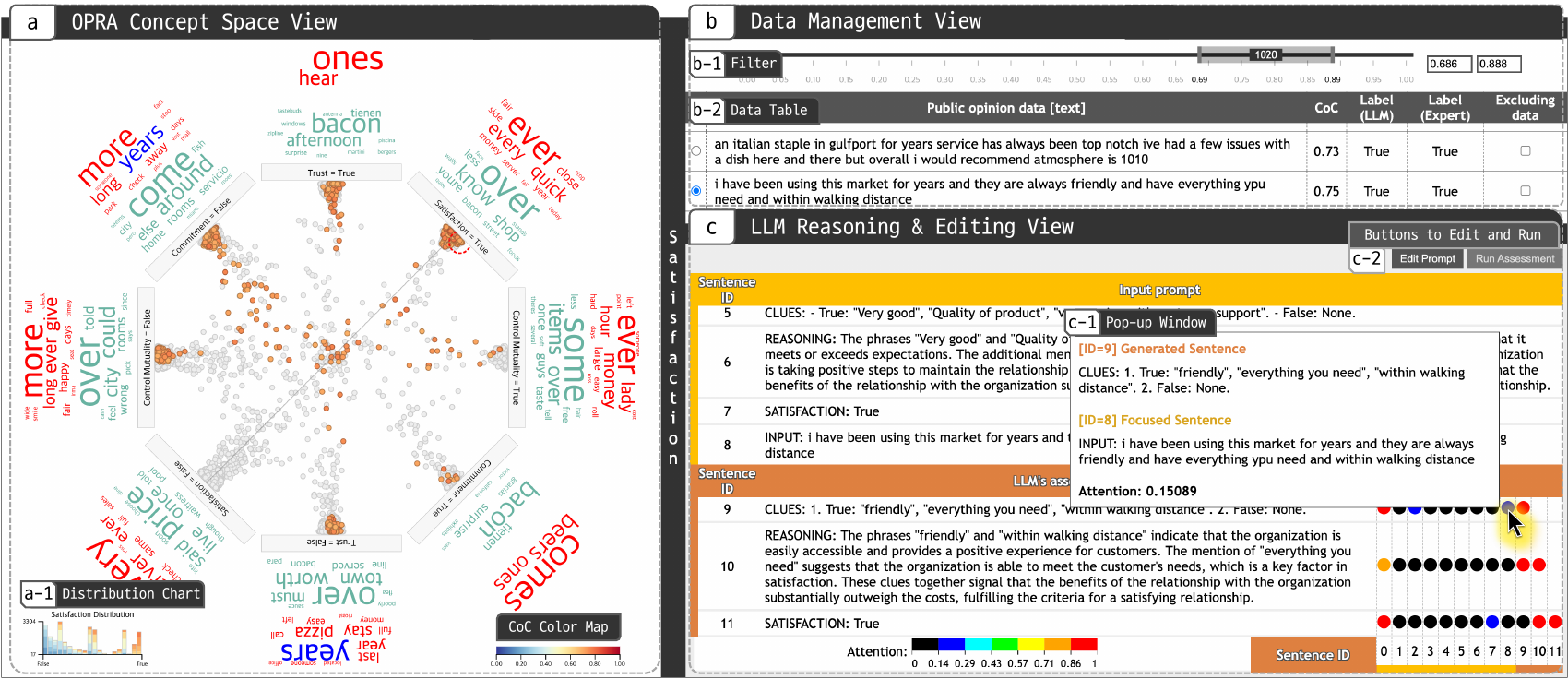}
    \caption{%
        The system overview of OPRA-Vis. OPRA-Vis consists of three tightly coordinated views. (a) is an OPRA Concept Space view to explore the certainty of concepts ($CoC$) relying on OPRA concepts using a scatter plot and provide the sentiment of words in sentences using tag clouds. (a-1) shows a distribution chart displaying data point density along the selected concept axis. (b) is a Data Management view. The Data Management view features a slider for data filtering, as presented in (b-1). The filter uses the $CoC$ range to update point visibility in the scatter plot in (a). The data table includes certainty, concept labels from LLM and experts, and data exclusion, as shown in (b-2). (c) is an LLM Reasoning \& Editing view for analyzing LLM's reasoning for labeling and editing prompts to adapt the user's knowledge. This view provides an attention visualization for analyzing the effectiveness of sentences in prompting. (c-1) is a pop-up window that organizes the generated sentences, focus sentences, and attention scores. The pop-up window is activated by hovering over the dot representing the sentence of attention visualization. (c-2) shows buttons to open the prompt editor and run the assessment using the edited prompt.
    }
    \label{fig:VA-LLM}
\end{figure*}

\subsection{Sentiment Classification}
\label{subsec:sentimet}
Figure~\ref{fig:flow_single} (b) illustrates the sentiment classification process. Sentences are normalized (lowercasing, lemmatization) and cleaned of non-informative tokens (stopwords, punctuation, emojis, very short tokens). They are then encoded using BERT$_\text{BASE}$ with a classification head, followed by a linear layer to yield sentiment probabilities (\textit{Positive} or \textit{Negative}). Based on these classifications, words are extracted from each sentence, and their frequencies are computed under the assigned \textit{Positive} or \textit{Negative} label. These results are then aggregated and organized by OPRA concept.

\subsection{Certainty of Concepts (CoC)}
\label{subsec:clarity}

While the concept labels yield outputs of \textit{True} or \textit{False}, mirroring PR expert labeling, public opinion sentences often contain ambiguities that complicate classification into clear-cut categories. To reflect these nuances in analysis, we define a concept-level certainty (certainty of concepts; $CoC$) that measures how strongly OPRA clues in a sentence support a given OPRA concept, as shown in Figure~\ref{fig:flow_single}~(c). Our system uses the $CoC$ to provide certainty-based filtering in the analysis of LLM's reasoning for OPRA concept labeling.

To compute the $CoC$ for OPRA-informed clues, we embed sentences using ${\text{BERT}}_{\text{BASE}}$, as shown in Figure~\ref{fig:flow_single}~(c). We apply dot-product attention between the sentence embedding and \textit{expert instructions} of each OPRA concept label (\textit{True} and \textit{False}). Dot-product attention measures semantic similarity between embeddings in the same vector space, enabling direct comparison without introducing additional parameters. Let $\mathbf{h}_{i} \in \mathbb{R}^{d_k}$ be the query embedding of the $i$-th sentence and $\mathbf{C}_j^{(\ell,k)} \in \mathbb{R}^{d_k}$ be the $k$-th key embedding for concept $j$ with label $\ell \in \{\text{true}, \text{false}\}$, where $d_k$ is the embedding dimension. We compute the certainty of concepts $CoC$ as follows.

\noindent
{%
\small
\begin{align}
    w_j^{(i,\ell)} &= \text{mean} \left( \max_k \left[ \text{softmax}\left( \mathbf{h}_i \cdot \left( \mathbf{C}_j^{(\ell,k)} \right)^T / \sqrt{d_k} \right) \cdot \left| \mathbf{C}_j^{(\ell, k)} \right| \right] \right) \notag \\
    CoC_{ij} &= \begin{cases}
        w_j^{(i,\text{true})} / \sum_\ell w_j^{(i,\ell)}, &\text{if } \ \sum_\ell w_j^{(i,\ell)} > 0 \\
        0.5, &\text{otherwise}
    \end{cases}
    \label{eq:coc}
\end{align}%
}%

\noindent
Here, the max operation selects the highest attention score within each \textit{expert instruction}, while the mean operation averages these maximum scores across multiple \textit{expert instructions} for the same concept-label pair. The computed $CoC$ values are standardized and rescaled to $[0,1]$ for consistent comparison and intuitive interpretation. These $CoC$ values serve as quantitative indicators of concept certainty, supporting analytical evaluation.
\section{OPRA-Vis: LLM-Assisted Visual Analytics System with CoT Prompting for OPRA}
\label{sec:system}
This section presents OPRA-Vis, a system designed to assist PR experts in OPRA labeling through three interconnected components (Figure~\ref{fig:VA-LLM}): (a) an OPRA Concept Space view, (b) a Data Management view, and (c) an LLM Reasoning \& Editing view. These components are tightly integrated, with interactions between (a) and (b) as well as between (b) and (c), enabling seamless exploration, analysis, and iterative editing.

\subsection{OPRA Concept Space View}
\label{subsec:swt_view}
\begin{figure}[t]
    \centering
    \includegraphics[width=\linewidth]{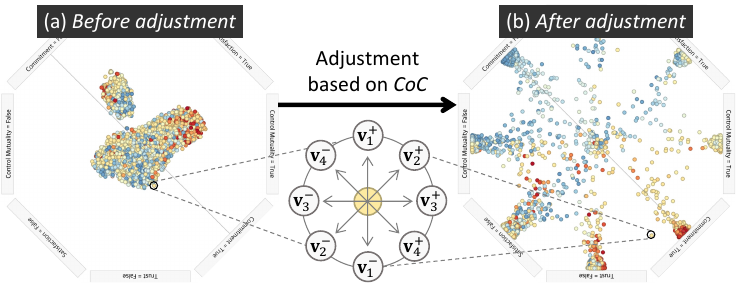}
    \caption{%
        Position adjustment using the gravity model weighted by $CoC$. (a) Before: Initial t-SNE layout of sentence embeddings forms dense clusters without clear concept alignment. (b) After: A gravity model uses \textit{True} and \textit{False} label vertices ($\mathbf{v}_j^{+}$ and $\mathbf{v}_j^{-}$ in Equation~\ref{eq:gravitational_adjustment}) as gravitational attractors, where each sentence is pulled toward its corresponding concept edge with force proportional to its $CoC$ deviation from 0.5. Sentences with $CoC$ values distant from 0.5 are pulled more strongly toward concept edges, while those near 0.5 experience weaker forces and remain closer to the center. This produces eight radial structures that separate concepts by certainty levels.
    }
    \label{fig:coc_adjustment}
\end{figure}
The OPRA Concept Space view, shown in Figure~\ref{fig:VA-LLM}~(a), visualizes public opinion texts according to the four OPRA concepts---\textit{Trust}, \textit{Satisfaction}, \textit{Commitment}, and \textit{Control Mutuality}---using a scatter plot and tag clouds. We adopt an octagonal layout that uses the angle to indicate the direction of the concept-specific \textit{True} or \textit{False} opposing label and the radius to encode $CoC$ in Equation~\ref{eq:coc}. Sentence embeddings are represented as points inside the octagon, while tag clouds are positioned around its perimeter. The embeddings are generated with $\text{BERT}_{\text{BASE}}$~\cite{devlin2018bert}, producing 768-dimensional vectors that are projected into 2D using t-SNE~\cite{tsne2008}. To better reflect OPRA-specific relationships, we further adjust the positions using Equation~\ref{eq:gravitational_adjustment}, which incorporates the $CoC$. We employ a physics-inspired gravity model~\cite{newton1833principia}, where the vertices representing OPRA concepts act as gravitational attractors. Sentence positions are iteratively updated based on their $CoC$ values through a constrained simulation.
\noindent%
{\normalsize%
\begin{align}%
    \mathbf{v}_{\text{target}}^{(i,j)} &= \mathbf{v}_j^{+} \cdot \textbf{1}_{CoC_{ij} > 0.5} + \mathbf{v}_j^{-} \cdot \mathbf{1}_{CoC_{ij} < 0.5} \notag \\
    \alpha_{ij} &= |CoC_{ij} - 0.5 | \cdot \alpha_{\text{base}} \cdot \mathbf{1}_{CoC_{ij} \neq 0.5} \notag \\
    F_{ij} &= G \frac {\alpha_{ij}} {(r_{ij} + \epsilon_1)^2}; \quad
    \textbf{F}_{ij} = F_{ij} \cdot \frac{\mathbf{v}_{\text{target}}^{(i,j)} - \mathbf{p}_i^{(t)}}{\left| \mathbf{v}_{\text{target}}^{(i,j)} - \mathbf{p}_i^{(t)} \right| + \epsilon_2} \notag \\
    \mathbf{u}_i^{(t+1)} &= \gamma \cdot \mathbf{u}_i^{(t)} + \delta \cdot \sum_j \mathbf{F}_{ij}; \quad
    \mathbf{p}_i^{(t+1)} = \mathbf{p}_i^{(t)} + \mathbf{u}_i^{(t+1)}
\label{eq:gravitational_adjustment}
\end{align}}%
\noindent%
Here, $\mathbf{v}_{j}^{+}$ and $\mathbf{v}_{j}^{-}$ represent \textit{True} and \textit{False} label vertices for concept $j$, $\mathbf{1}_{\{\cdot\}}$ is the indicator function, $\alpha_{\text{base}}$ is the base gravitational strength, $G$ is the gravitational constant, $r_{ij} = \|\mathbf{v}_{\text{target}}^{(i,j)} - \mathbf{p}_i^{(t)}\|_2$ is the Euclidean distance, $\mathbf{u}_i^{(t)}$ is the velocity vector, $\gamma$ is the damping factor, $\delta$ is the step size, $\epsilon_1$ prevents force divergence when $r_{ij} = 0$, and $\epsilon_2$ prevents division by zero when computing directions. In our implementation, we use $\alpha_{\text{base}}=2.0$, $G=1.0$, $\gamma=0.8$, $\delta=0.1$, $\epsilon_1=0.01$, and $\epsilon_2=10^{-10}$. These parameters were empirically selected and can be adjusted. We will explore systematic optimization methods in future work. The gravitational force follows an inverse square law, where $\alpha_{ij}$ acts as the gravitational mass proportional to the deviation of $CoC_{ij}$ from 0.5. When $CoC_{ij}~=~0.5$, no gravitational force is applied as the sentence shows intermediate $CoC$ values. Positions are constrained within a unit circle through normalization to maintain visual coherence. This iterative gravitational simulation continues for up to 200 iterations or until position changes fall below $10^{-4}$, creating spatial clustering where sentences are gravitationally attracted toward vertices corresponding to their respective concept labels. Figure~\ref{fig:coc_adjustment} illustrates this transformation: (a) shows the initial t-SNE layout, and (b) shows the adjusted plot reflecting $CoC$ values with gravitational attractors.

\begin{figure}[t]
    \centering
    \includegraphics[width=\linewidth]{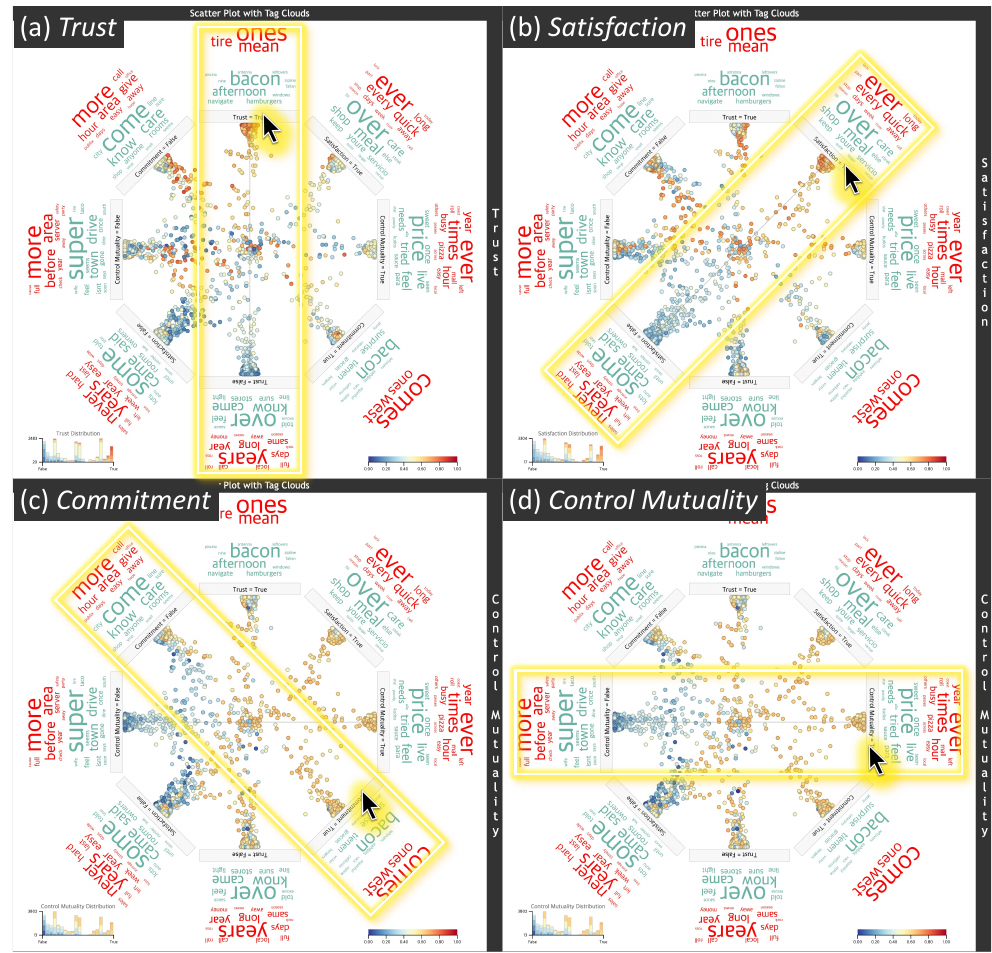}
    \caption{%
        Interaction of the OPRA Concept Space view. Selecting an edge of the octagonal layout activates the corresponding OPRA concept; the scatter plot then color-encodes the $CoC$ (Section~\ref{subsec:clarity}) for that concept to provide an overview of how opinions are distributed. (a) to (d) show the views with \textit{Trust}, \textit{Satisfaction}, \textit{Commitment}, and \textit{Control Mutuality} selected, respectively.
    }
    \label{fig:vis_scatter_tag}
\end{figure}

The scatter plot enables interactive concept selection. Clicking on an OPRA concept label on the concept edge updates the color encoding based on the CoC values for that concept using a blue-yellow-red color scheme. Figures~\ref{fig:vis_scatter_tag}~(a)--(d) illustrate scatter plots for the cases when the user selects \textit{Trust}, \textit{Satisfaction}, \textit{Commitment}, and \textit{Control Mutuality}, respectively.

To illustrate how $CoC$ values translate into positions along the concept axis, we provide a distribution chart for the user-selected OPRA concept label from the scatter plot, as illustrated in Figure~\ref{fig:VA-LLM}~(a-1). The axis spans from \textit{False} to \textit{True}, and each data point is orthogonally projected onto the axis, yielding a position within $[0,1]$. The distribution is shown as a histogram with bars stacked by $CoC$ values and a $\log_{10}$-scaled y-axis to emphasize patterns across frequency ranges. Users can toggle between linear, $\ln$, $\log_2$, and $\log_{10}$ scales by clicking on the chart title, enabling flexible analysis of both sparse and dense regions of the position distribution.

To facilitate analysis of LLM's reasoning for labeling, we visualize sentiment related to OPRA concepts using tag clouds surrounding the scatter plot. These tag clouds are generated by extracting word frequencies from sentiment-encoded sentences, with word size proportional to frequency and color indicating sentiment---red for \textit{Negative} and green for \textit{Positive}. Highlighted words correspond to those in the sentences selected from Figure~\ref{fig:VA-LLM}~(b), supporting concept-specific reasoning.

\subsection{Data Management View}
\label{subsec:dm_view}
The Data Management view, shown in Figure~\ref{fig:VA-LLM}~(b), supports usability and reduces human workload. It consists of a filter (b-1) and a data table (b-2). The filter adjusts the $CoC$ range, rendering out-of-range points semi-transparent in the scatter plot (a) and hiding them from the table. The table displays sentences with $CoC$, human-coded ground truth, and LLM assessments. For a selected sentence, the clues and step-by-step reasoning for labeling are interactively linked to the view in (c). The filter identifies a subset using $CoC$, and the following operations are executed in views (a), (b), and (c) on this subset.

Before analysis, the system simplifies the dataset by removing redundant or uninformative sentences, identified via cosine distance over sentence embeddings. This reduction lowers re-evaluation costs and manual review effort. Uninformative content, such as emojis or nonsensical text (e.g., \textit{yh}, \textit{hello}, \textit{aaaaa}, \textit{qwer}), obscures OPRA concepts or adds little semantic value. The data table then supports prompt adjustment with ground-truth labels, enhancing usability by facilitating effortless problem identification and data management.

\subsection{LLM Reasoning \& Editing View}
\label{subsec:llm_view}
The LLM Reasoning \& Editing view in Figure~\ref{fig:VA-LLM}~(c) provides an attention visualization to examine LLM's reasoning for labeling and to support prompt modification.

\begin{figure}[t]
    \centering
    \includegraphics[width=.9\linewidth]{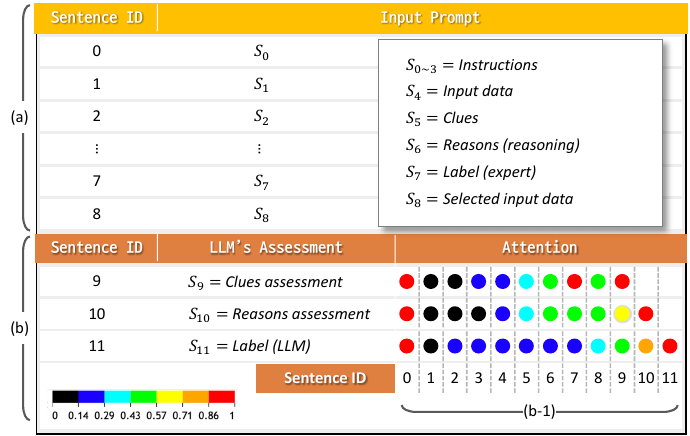}
    \caption{%
        Attention visualization in LLM Reasoning \& Editing view. (a) shows the LLM input prompts divided into $S_{0\textrm{--}8}$ according to \textit{Sentence ID}. (b) shows $S_{9\textrm{--}11}$, the sentences generated by LLM, and attention visualization. (b-1) is an attention visualization to identify sentences influenced by sentences generated from LLM's assessment.
    }
    \label{fig:LLM_reasoning}
\end{figure}

\begin{figure*}[t]
    \centering
    \includegraphics[width=\linewidth]{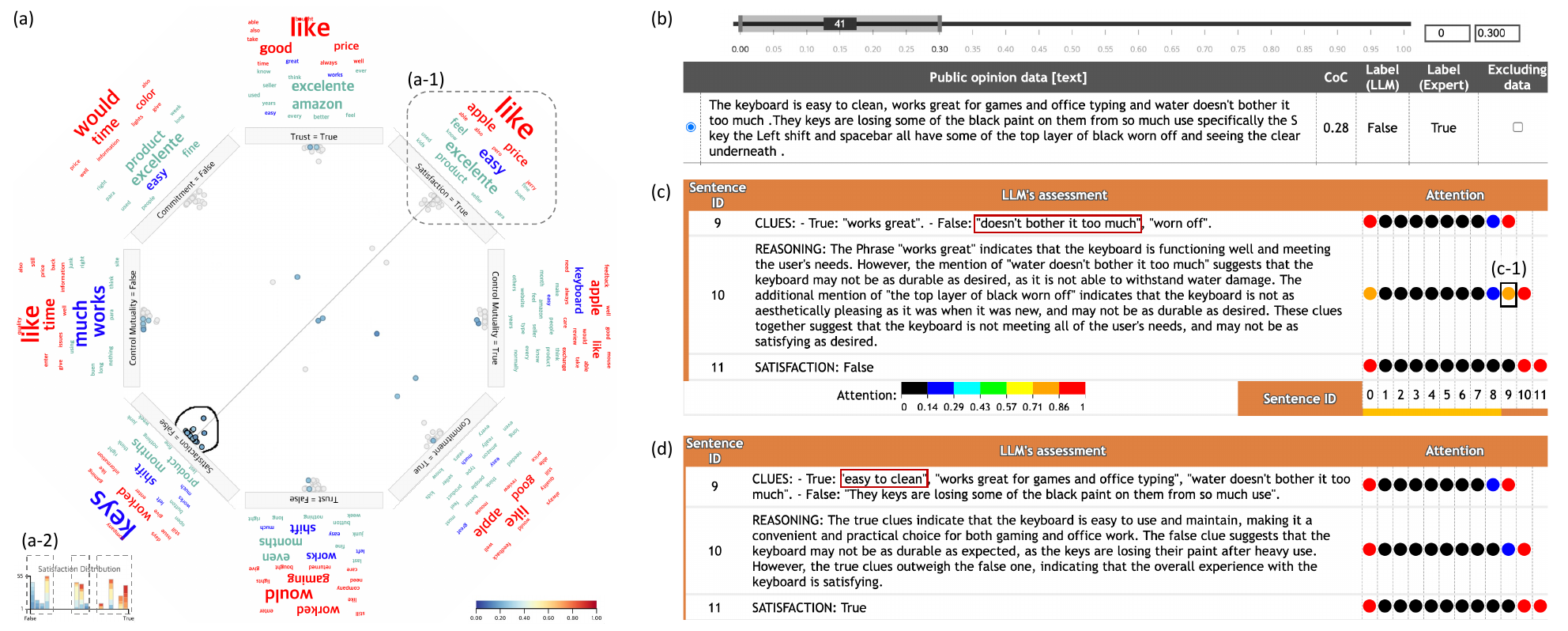}
    \caption{%
        LLM's reasoning for labeling on Amazon product review data. (a) shows the OPRA Scpace view with \textit{Satisfaction} selected by the user. b) presents public opinion data where LLM and experts label differently. (c) reveals that LLM has a flawed reasoning for labeling. (d) is a prompt where the user revises the clues and displays the assessment results, which show that LLM's reasoning for labeling has been revised.
    }
    \label{fig:case_amazon}
\end{figure*}

Figure~\ref{fig:LLM_reasoning} details the LLM Reasoning \& Editing view. In (a), $S_{0\textrm{--}8}$ represent the LLM input prompts. Within this prompt structure, $S_{0\textrm{--}3}$ serve as task instructions for OPRA, while $S_{4\textrm{--}7}$ constitute an \textit{FSE} ($k=1$), as described in Section~\ref{subsec:polarity_sentiment}. $S_4$ is an example input sentence, and $S_5$ provides the corresponding clues extracted from $S_4$. $S_6$ shows the reasoning process based on $S_4$ \& $S_5$, and $S_7$ provides the final assessment decision. $S_8$ is a selected sentence from the data table in Figure~\ref{fig:VA-LLM} (b-2), which serves as the target input for the assessment. Based on this prompt, the LLM infers the label of $S_8$. In Figure~\ref{fig:LLM_reasoning}~(b), the sentences $S_{9\textrm{--}11}$ are generated by the LLM according to the established reasoning pattern. $S_9$ identifies clues in $S_8$, and $S_{10}$ provides reasoning based on the clues identified in $S_9$. $S_{11}$ presents the final label.

Figure~\ref{fig:LLM_reasoning}~(b-1) visualizes inter-sentence attention (ISA)~\cite{seo2025sentence}, highlighting how the LLM utilizes preceding context. The complexity of the transformer architecture makes it unclear which elements have the greatest influence on generation. $\text{ISA}(S_a,S_b)$ represents the aggregated token-level attention from $S_a$ to $S_b$, and the color on each sentence dot indicates incoming attention, demonstrating how $S_{9\text{--}11}$ were influenced. In (b), $S_{10}$ is primarily influenced by $S_0$, which provides task instructions (e.g., \textit{``This is a satisfaction assessment tool for evaluating reviews.''}). The next most influential is $S_9$, which is the clue sentence derived from $S_8$. These influences suggest that $S_{10}$ was generated appropriately.

Figure~\ref{fig:VA-LLM}~(c-1) shows a pop-up window triggered by hovering over a dot in the attention visualization. It presents the generated sentence, the focused sentence, and the attention score. This view supports analysis of sentence-level influences on LLM outputs, providing insight into its reasoning process. (c) enables prompt editing based on this analysis. (c-2) includes buttons for opening the prompt editor and rerunning the LLM, with results updated in the data table in Figure~\ref{fig:VA-LLM}~(b-2).
\section{Use Scenarios}
\label{sec:case_study}
This section presents two real-world use scenarios demonstrating OPRA-Vis's labeling and reasoning capabilities using Amazon~\cite{amazon} and Google Local~\cite{li2022uctopic} reviews.

\begin{figure}[t]
    \centering
    \includegraphics[width=\linewidth]{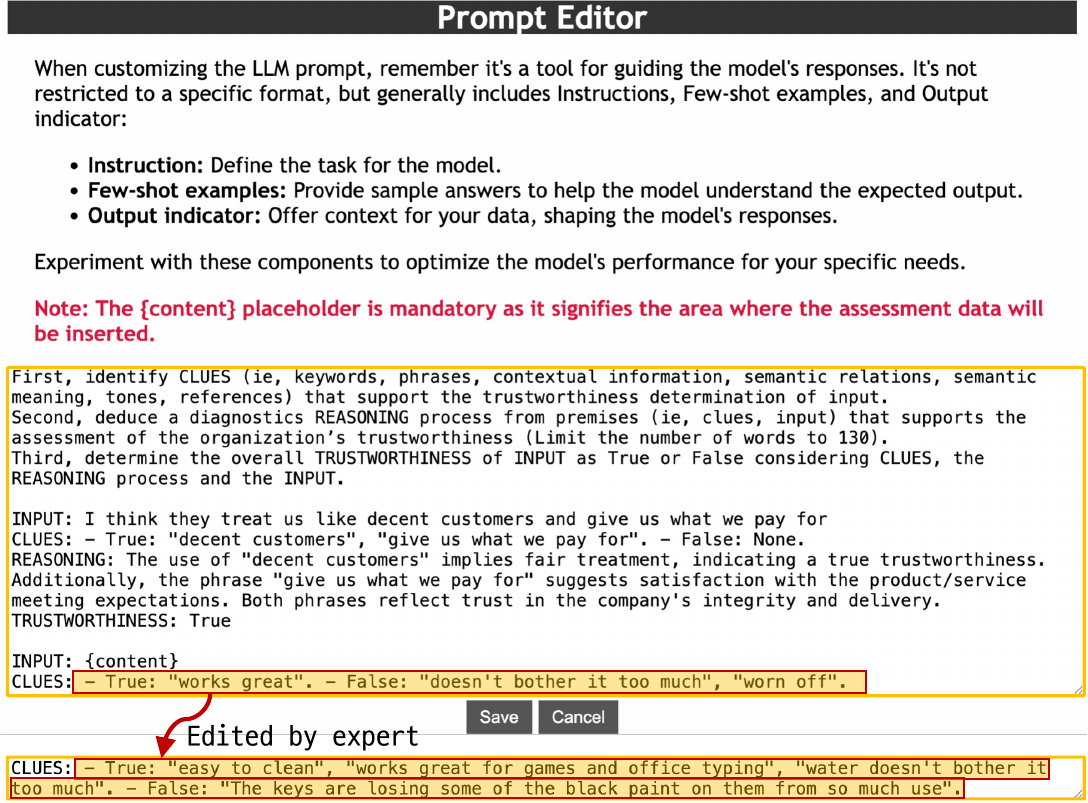}
    \caption{%
        Prompt editor. Text highlighted in \colorbox{lightorange}{orange} shows the prompt before and after editing. The prompt in the blue edit box is freely editable.
    }
    \label{fig:edit_prompt}
\end{figure}

\subsection{Linguistic Negation Correction}
\label{subsec:case_amazon}
This use scenario involves 198 Amazon product reviews~\cite{amazon}. The OPRA concept labels as the ground-truth of these reviews are labeled by two PR researchers. To address LLM misclassifications, our system enables analyst-driven decision-making through prompt editing. This includes certainty-based filtering, sentiment-aware tag clouds, and attention visualization. Figure~\ref{fig:case_amazon} provides an example of the relabeling process.

Guided by the distribution chart in Figure~\ref{fig:case_amazon}~(a), we partition the data into three $CoC$ groups, as shown within the three gray dotted frames. While exploring the group (a-2) with $CoC$ in the range of 0 to 0.3 using the filtering on the Data Management view in (b), we capture that 41 sentences surface several mismatches between expert and LLM labels. (b) shows a representative case, \textit{``The keyboard is easy to clean, [...] the clear underneath''}. The expert labels this sentence \textit{True} for \textit{Satisfaction}, as indicated by \textit{``Label (Expert) = True''}, whereas the LLM labels it \textit{False}. As we examine the LLM's reasoning for labeling in (c), we find that linguistic negation underlies this error. Negation is a well-known source of ambiguity in LLM's labeling tasks~\cite{garcia2023dataset}.

\begin{figure*}[t]
    \centering
    \includegraphics[width=\linewidth]{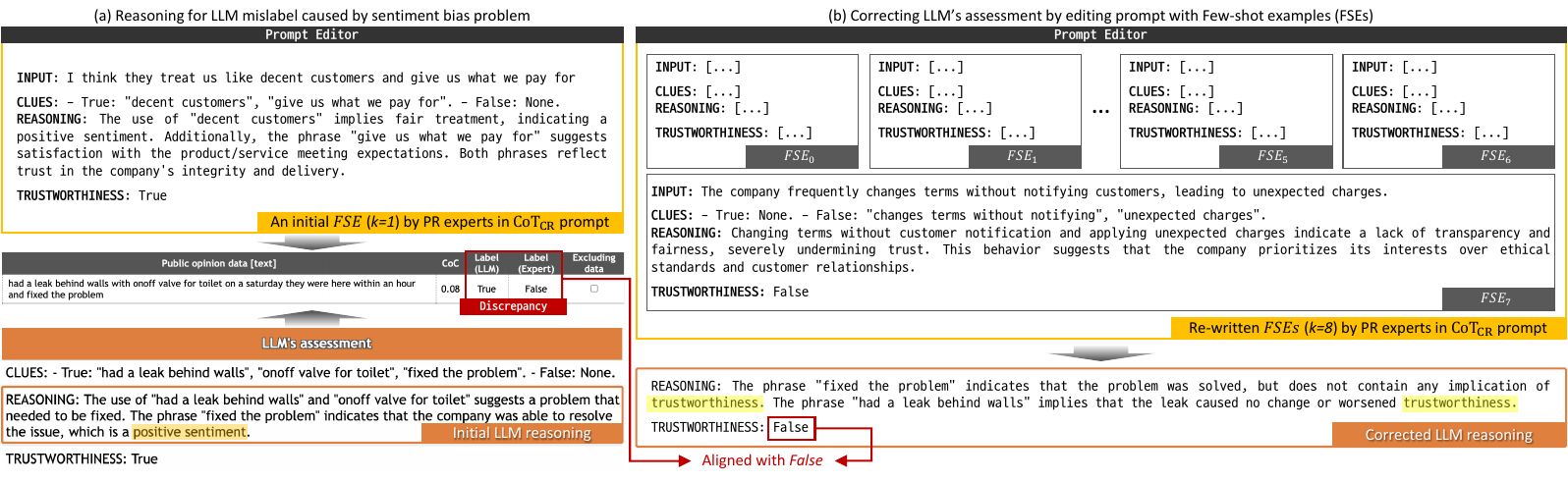}
    \caption{%
        Correcting a sentiment-biased mislabel. (a) The LLM incorrectly labels the OPRA concept \textit{Trust} as \textit{True} by focusing on positive-sentiment clues, such as \textit{``fixed the problem''}. (b) After PR experts rewrite few-shot examples (\textit{FSEs}) in the Prompt Editor to connect \textit{Trust} with cues of transparency and fairness, the LLM's reasoning and assessment align with the expert label of \textit{False}.
    }
    \label{fig:case_google}
\end{figure*}

In Figure~\ref{fig:case_amazon}~(c), LLM misclassified the sentence \textit{``water doesn't bother it too much''}, which conveys a \textit{True} meaning, as \textit{False}. This is shown in \texttt{CLUES} (\textit{Sentence~ID}$=$9), where LLM focuses on the phrase \textit{``doesn't bother it too much''}, highlighting its difficulty in handling negation. Consequently, LLM inferred the \texttt{SATISFACTION} label (\textit{Sentence~ID}$=$11) as \textit{False}, based on \texttt{REASONING} (\textit{Sentence~ID}$=$10). (c-1) visualizes that \texttt{REASONING} (\textit{Sentence~ID}$=$10) was most influenced by \texttt{CLUES} (\textit{Sentence~ID}$=$9), shown as an orange dot. For clarity, we excluded \textit{Sentence~ID}$=$0 and \textit{Sentence~ID}$=$10, both represented as red dots with the highest attention scores, from this inter-sentence attention analysis. \textit{Sentence~ID}$=$0 corresponds to \textit{Instructions} and appears in nearly all generations, making it unsuitable for analyzing sentence-specific influence (see Section~\ref{subsec:llm_view}). \textit{Sentence~ID}$=$10 was also excluded as it is self-referential in the attention map. These findings guided prompt editing, as shown in Figure~\ref{fig:edit_prompt}.

We revise the prompt as shown in Figure~\ref{fig:edit_prompt}, where yellow highlights indicate the edited text. Specifically, we change \textit{``water doesn't bother it too much''} from \textit{False} to \textit{True}. The phrase \textit{``easy to clean''} includes the word \textit{``easy''}, which suggests a \textit{True} connotation and appears blue in the tag cloud for ``\textit{Satisfaction=True}'' (Figure~\ref{fig:case_amazon}~(a-1)). Therefore, this clue is retained. We also add \textit{``works great for games and office typing''} to the \textit{True} category in \texttt{CLUES}, as it supports \textit{Satisfaction}. Although only the blue-highlighted text is revised in this case, any part of the prompt remains editable.

Figure~\ref{fig:case_amazon}~(d) shows the updated model assessment following the edited prompt. 
The \texttt{CLUES} at \textit{Sentence~ID}=9 reflects the revised input, while the \texttt{REASONING} at \textit{Sentence~ID}=10 is newly generated based on these modified clues. Influenced by this reasoning, LLM corrects its assessment of \textit{Satisfaction} to \textit{True}. OPRA-Vis propagates the edit by updating the shared prompt template and the exemplar pool. Then it reassesses the entire dataset, updating labels and rationales. This process clarifies how the prompt manages phenomena like negation, updates similar cases, and reduces inconsistencies between the LLM and expert labels.

\subsection{Correcting Mislabeling Caused by Sentiment Bias}
\label{subsec:case_google}
This scenario, illustrated in Figure~\ref{fig:case_google}, highlights a failure in concept labeling due to sentiment-bias reasoning and its correction through few-shot learning. The LLM confuses overall positivity with OPRA-informed clues, leading to the mislabeling of the sentence. We demonstrate this scenario using Google Local reviews~\cite{li2022uctopic}, which contains 666 million reviews of U.S. businesses as of September 2021. The dataset comprises ratings, textual content, images, and metadata, including location, category, and pricing information.

Figure~\ref{fig:case_google} (a) displays the initial mislabeling. In a sentence from public opinion data indicating that the issue was resolved, the LLM focuses on positive-sentiment phrases, such as \textit{``fixed the problem''}. It interprets these as evidence for the OPRA concept \textit{Trust} and outputs \textit{True}, while the expert label is \textit{False}. This error reveals a reasoning shortcut: the LLM equates problem resolution or satisfaction with trustworthiness, rather than evaluating OPRA-informed clues such as transparency, fairness, and ethical conduct. The table emphasizes the discrepancy between \textit{``Label~(LLM)~=~True''} and \textit{``Label~(Expert)~=~False''}, which aligns with sentiment-driven reasoning, as indicated by the orange-highlighted \textit{``\colorbox{lightorange}{positive sentiment}''} in the initial LLM reasoning. Figure~\ref{fig:case_google} (b) shows the correction made after PR experts rewrite and expand the \textit{FSEs} in the Prompt Editor to anchor better \textit{Trust} to the OPRA-informed clues, as described in Section~\ref{subsec:opra}. In the corrected reasoning, the LLM explicitly dismisses positive tone as sufficient evidence and instead grounds its assessment on \textit{``\colorbox{lightyellow}{trustworthiness}''} in the \texttt{REASONING} text. This assessment aligns with the expert label of \textit{False}, demonstrating that editing \textit{FSEs} to align with concepts effectively mitigates sentiment bias and restores construct-centered labeling.
\section{Quantitative Evaluation}
This section presents quantitative evaluations of OPRA-Vis, focusing on labeling accuracy and the effectiveness of prompts. We utilize expert-labeled datasets that include 198 Amazon product reviews (\textit{Amazon}), 100 Google Local reviews (\textit{Google}), and 100 Internet Movie Database reviews (\textit{IMDB})~\cite{maas2011learning}. Each dataset is annotated with OPRA concept labels, creating a controlled setting to compare performance across various PR domains.

\subsection{Model selection via assessment accuracy}
\label{sec:quantitative_evaluation}
We evaluate the core model integrated into OPRA-Vis by comparing three well-established PLMs: Llama 2 (7B)~\cite{touvron2023llama}, GPT-3~\cite{brown2020language}, and Gemma (7B)~\cite{team2024gemma}. The goal of this comparison is not to establish SOTA performance, but to identify a suitable base model for OPRA-Vis within our development constraints. This ensures that the evaluation is procedural and reproducible rather than tied to the specific vintage of any given model. To ensure consistency across models, we applied CoT prompting with embedded clues and reasoning to incorporate PR theory into the evaluation process. Table~\ref{table:quantitative_evaluation} reports the labeling accuracy of concepts for each evaluated model.
\begin{table}[t]
    \footnotesize
    \caption{
        Comparisons of OPRA concept labeling accuracy. We evaluate Llama 2 (7B), GPT-3, and Gemma (7B) on three datasets: Amazon product reviews (\textit{Amazon}), Google Local reviews (\textit{Google}), and Internet Movie Database reviews (\textit{IMDB}). Abbreviations: \textit{Satis.} = Satisfaction, \textit{Commit.} = Commitment, and \textit{C.M.} = Control Mutuality.
    }
    \centering
    \label{table:quantitative_evaluation}
    \begin{tabular}{p{0.05\textwidth}p{0.05\textwidth}p{0.1\textwidth}p{0.1\textwidth}p{0.1\textwidth}p{0.1\textwidth}p{0.1\textwidth}}
        \toprule
            \multirow{2}{*}{\textbf{Dataset}} & \multicolumn{1}{c}{\multirow{2}{*}{\textbf{Model}}} & \multicolumn{4}{c}{\textbf{OPRA Labeling Accuracy}} & \multicolumn{1}{c}{\textbf{Average}}\\ 
            & & \multicolumn{1}{c}{\textit{Trust}} & \multicolumn{1}{c}{\textit{Satis.}} & \multicolumn{1}{c}{\textit{Commit.}} & \multicolumn{1}{c}{\textit{C.M.}} & \multicolumn{1}{c}{\textbf{Accuracy}} \\
        \midrule
            \multicolumn{7}{c}{\underline{One-shot settings}} \\
        \midrule
            \multirow{3}{*}{\textit{Amazon}} &
            \multicolumn{1}{c}{Llama 2} & \multicolumn{1}{r}{70\%} & \multicolumn{1}{r}{79\%} & \multicolumn{1}{r}{49\%} & \multicolumn{1}{r}{44\%} & \multicolumn{1}{r}{60.50\%} \\ \cline{2-7} 
            & \multicolumn{1}{c}{GPT-3} &
            \multicolumn{1}{r}{\textbf{79\%}} & \multicolumn{1}{r}{\textbf{90\%}} & \multicolumn{1}{r}{55\%} & \multicolumn{1}{r}{54\%} & \multicolumn{1}{r}{69.50\%} \\ \cline{2-7}
            & \multicolumn{1}{c}{Gemma} &
            \multicolumn{1}{r}{77\%} & \multicolumn{1}{r}{88\%} & \multicolumn{1}{r}{\textbf{63\%}} & \multicolumn{1}{r}{\textbf{72\%}} & \multicolumn{1}{r}{\textbf{75\%}} \\ 
        \midrule
            \multirow{3}{*}{\textit{Google}} &
            \multicolumn{1}{c}{Llama 2} &
            \multicolumn{1}{r}{66\%} & \multicolumn{1}{r}{70\%} & \multicolumn{1}{r}{49\%} & \multicolumn{1}{r}{53\%} & \multicolumn{1}{r}{59.50\%} \\ \cline{2-7}
            & \multicolumn{1}{c}{GPT-3} &
            \multicolumn{1}{r}{\textbf{76\%}} & \multicolumn{1}{r}{\textbf{84\%}} & \multicolumn{1}{r}{68\%} & \multicolumn{1}{r}{49\%} & \multicolumn{1}{r}{69.25\%} \\ \cline{2-7}
            & \multicolumn{1}{c}{Gemma} &
            \multicolumn{1}{r}{\textbf{76\%}} & \multicolumn{1}{r}{83\%} & \multicolumn{1}{r}{\textbf{72\%}} & \multicolumn{1}{r}{\textbf{55\%}} & \multicolumn{1}{r}{\textbf{71.50\%}} \\
        \midrule
            \multirow{3}{*}{\textit{IMDB}} &
            \multicolumn{1}{c}{Llama 2} &
            \multicolumn{1}{r}{59\%} & \multicolumn{1}{r}{70\%} & \multicolumn{1}{r}{58\%} & \multicolumn{1}{r}{49\%} & \multicolumn{1}{r}{59\%} \\ \cline{2-7}
            & \multicolumn{1}{c}{GPT-3} &
            \multicolumn{1}{r}{73\%} & \multicolumn{1}{r}{\textbf{92\%}} & \multicolumn{1}{r}{72\%} & \multicolumn{1}{r}{46\%} & \multicolumn{1}{r}{70.75\%} \\ \cline{2-7}
            & \multicolumn{1}{c}{Gemma} &
            \multicolumn{1}{r}{\textbf{75\%}} & \multicolumn{1}{r}{\textbf{92\%}} & \multicolumn{1}{r}{\textbf{81\%}} & \multicolumn{1}{r}{\textbf{64\%}} & \multicolumn{1}{r}{\textbf{78\%}} \\
        \midrule
        \multicolumn{7}{c}{\underline{Few-shot ($k=8$) settings}} \\ 
        \midrule
            \multirow{3}{*}{\textit{Amazon}} &
            \multicolumn{1}{c}{Llama 2} & \multicolumn{1}{r}{67\%} & \multicolumn{1}{r}{84\%} & \multicolumn{1}{r}{54\%} & \multicolumn{1}{r}{46\%} & \multicolumn{1}{r}{62.75\%} \\ \cline{2-7} 
            & \multicolumn{1}{c}{GPT-3} &
            \multicolumn{1}{r}{\textbf{78\%}} & \multicolumn{1}{r}{\textbf{89\%}} & \multicolumn{1}{r}{58\%} & \multicolumn{1}{r}{55\%} & \multicolumn{1}{r}{70\%} \\ \cline{2-7}
            & \multicolumn{1}{c}{Gemma} &
            \multicolumn{1}{r}{74\%} & \multicolumn{1}{r}{\textbf{89\%}} & \multicolumn{1}{r}{\textbf{60\%}} & \multicolumn{1}{r}{\textbf{87\%}} & \multicolumn{1}{r}{\textbf{77.50\%}} \\ 
        \midrule
            \multirow{3}{*}{\textit{Google}} &
            \multicolumn{1}{c}{Llama 2} &
            \multicolumn{1}{r}{70\%} & \multicolumn{1}{r}{84\%} & \multicolumn{1}{r}{65\%} & \multicolumn{1}{r}{43\%} & \multicolumn{1}{r}{65.50\%} \\ \cline{2-7}
            & \multicolumn{1}{c}{GPT-3} &
            \multicolumn{1}{r}{\textbf{76\%}} & \multicolumn{1}{r}{\textbf{85\%}} & \multicolumn{1}{r}{\textbf{72\%}} & \multicolumn{1}{r}{46\%} & \multicolumn{1}{r}{69.75\%} \\ \cline{2-7}
            & \multicolumn{1}{c}{Gemma} &
            \multicolumn{1}{r}{71\%} & \multicolumn{1}{r}{84\%} & \multicolumn{1}{r}{70\%} & \multicolumn{1}{r}{\textbf{78\%}} & \multicolumn{1}{r}{\textbf{75.75\%}} \\
        \midrule
            \multirow{3}{*}{\textit{IMDB}} &
            \multicolumn{1}{c}{Llama 2} &
            \multicolumn{1}{r}{64\%} & \multicolumn{1}{r}{83\%} & \multicolumn{1}{r}{63\%} & \multicolumn{1}{r}{46\%} & \multicolumn{1}{r}{64\%} \\ \cline{2-7}
            & \multicolumn{1}{c}{GPT-3} &
            \multicolumn{1}{r}{\textbf{73\%}} & \multicolumn{1}{r}{\textbf{94\%}} & \multicolumn{1}{r}{69\%} & \multicolumn{1}{r}{46\%} & \multicolumn{1}{r}{70.50\%} \\ \cline{2-7}
            & \multicolumn{1}{c}{Gemma} &
            \multicolumn{1}{r}{71\%} & \multicolumn{1}{r}{88\%} & \multicolumn{1}{r}{\textbf{89\%}} & \multicolumn{1}{r}{\textbf{97\%}} & \multicolumn{1}{r}{\textbf{86.25\%}} \\
        \bottomrule
    \end{tabular}
\end{table}
The average accuracy of Gemma surpassed that of the other PLMs in both one-shot and few-shot ($k=8$) scenarios. Across all datasets, Gemma achieved the highest overall average, followed by GPT-3 and Llama 2. All models showed improved accuracy under few-shot conditions compared to one-shot, except for GPT-3, which exhibited a slight decrease in accuracy for the \textit{Google} dataset. Examining individual OPRA concepts, GPT-3 achieved 1--6\% higher accuracy than Gemma for \textit{Trust} and \textit{Satisfaction} in the one-shot \textit{Amazon} dataset, as well as in few-shot evaluations across \textit{Amazon}, \textit{Google}, and \textit{IMDB}. However, Gemma consistently outperformed GPT-3 for \textit{Commitment} and \textit{Control Mutuality} across most scenarios. These results indicate that OPRA-Vis effectively integrates PR theory by utilizing Gemma as its core model.

\subsection{Prompting Test}
\label{subsec:ablation_test}
Text classification performance varies by prompting method, as shown in Table~\ref{table:albation_test}. To determine the most effective approach grounded in PR theory, we compare LLM labeling accuracy against expert-annotated ground truth across four OPRA concepts using three prompting strategies: vanilla prompting (Vanilla), CoT prompting (CoT), and CoT with clues and reasoning (CoT$_\text{CR}$), evaluated under one-shot and few-shot settings. Among the three methods, CoT$_\text{CR}$ consistently achieves the highest accuracy across datasets and concepts, with notable improvements in challenging categories such as \textit{Control Mutuality}. For example, on the \textit{IMDB} dataset under the one-shot setting, accuracy rises from 18\% (Vanilla) to 64\% (CoT$_\text{CR}$). While CoT provides moderate gains, adding expert-informed clues and reasoning in CoT$_\text{CR}$ significantly reduces the gap between LLM predictions and human labels. This method also demonstrates stable, robust performance in the few-shot ($k=8$) setting, highlighting the importance of structured domain knowledge for aligning LLM outputs with expert judgment. Consequently, CoT$_\text{CR}$ is adopted as the default prompting strategy in OPRA-Vis.

\begin{table}[t]
    \footnotesize
    \caption{
        Prompting test for employing expert knowledge across different settings. We compare vanilla prompting (Vanilla), CoT prompting (CoT), and CoT prompting with clues and reasons (CoT$_\text{CR}$). The evaluation uses Amazon product reviews (\textit{Amazon}), Google local reviews (\textit{Google}), and Internet Movie Database reviews (\textit{IMDB}).
    }
    \centering
    \label{table:albation_test}
    \begin{tabular}{p{0.05\textwidth}p{0.05\textwidth}p{0.1\textwidth}p{0.1\textwidth}p{0.1\textwidth}p{0.1\textwidth}p{0.1\textwidth}}
        \toprule
            \multirow{2}{*}{\textbf{Dataset}} & \multirow{2}{*}{\textbf{Method}} & \multicolumn{4}{c}{\textbf{OPRA Labeling Accuracy}} & \multicolumn{1}{c}{\textbf{Average}}\\ 
            & & \multicolumn{1}{c}{\textit{Trust}} & \multicolumn{1}{c}{\textit{Satis.}} & \multicolumn{1}{c}{\textit{Commit.}} & \multicolumn{1}{c}{\textit{C.M.}} & \multicolumn{1}{c}{\textbf{Accuracy}} \\
        \midrule
            \multicolumn{7}{c}{\underline{One-shot settings}} \\
        \midrule
            \multirow{3}{*}{\textit{Amazon}} & \multicolumn{1}{c}{Vanilla} & \multicolumn{1}{r}{71\%} & \multicolumn{1}{r}{83\%} & \multicolumn{1}{r}{58\%} & \multicolumn{1}{r}{25\%} & \multicolumn{1}{r}{59.25\%} \\ \cline{2-7}
            & \multicolumn{1}{c}{CoT} & \multicolumn{1}{r}{74\%} & \multicolumn{1}{r}{86\%} & \multicolumn{1}{r}{51\%} & \multicolumn{1}{r}{43\%} & \multicolumn{1}{r}{63.50\%} \\ \cline{2-7}
            & \multicolumn{1}{c}{CoT$_\text{CR}$} & \multicolumn{1}{r}{\textbf{77\%}} & \multicolumn{1}{r}{\textbf{88\%}} & \multicolumn{1}{r}{\textbf{63\%}} & \multicolumn{1}{r}{\textbf{72\%}} & \multicolumn{1}{r}{\textbf{75\%}} \\ 
        \midrule
            \multirow{3}{*}{\textit{Google}} & \multicolumn{1}{c}{Vanilla} & \multicolumn{1}{r}{70\%} & \multicolumn{1}{r}{79\%} & \multicolumn{1}{r}{66\%} & \multicolumn{1}{r}{34\%} & \multicolumn{1}{r}{62.25\%} \\\cline{2-7}
            & \multicolumn{1}{c}{CoT} & \multicolumn{1}{r}{74\%} & \multicolumn{1}{r}{\textbf{83\%}} & \multicolumn{1}{r}{70\%} & \multicolumn{1}{r}{39\%} & \multicolumn{1}{r}{66.50\%} \\ \cline{2-7}
            & \multicolumn{1}{c}{CoT$_\text{CR}$} & \multicolumn{1}{r}{\textbf{76\%}} & \multicolumn{1}{r}{\textbf{83\%}} & \multicolumn{1}{r}{\textbf{72\%}} & \multicolumn{1}{r}{\textbf{55\%}} & \multicolumn{1}{r}{\textbf{71.50\%}} \\         
        \midrule
            \multirow{3}{*}{\textit{IMDB}} & \multicolumn{1}{c}{Vanilla} & \multicolumn{1}{r}{59\%} & \multicolumn{1}{r}{83\%} & \multicolumn{1}{r}{74\%} & \multicolumn{1}{r}{18\%} & \multicolumn{1}{r}{58.50\%} \\ \cline{2-7}
            & \multicolumn{1}{c}{CoT} & \multicolumn{1}{r}{74\%} & \multicolumn{1}{r}{88\%} & \multicolumn{1}{r}{\textbf{84\%}} & \multicolumn{1}{r}{32\%} & \multicolumn{1}{r}{69.50\%} \\ \cline{2-7}
            & \multicolumn{1}{c}{CoT$_\text{CR}$} & \multicolumn{1}{r}{\textbf{75\%}} & \multicolumn{1}{r}{\textbf{92\%}} & \multicolumn{1}{r}{81\%} & \multicolumn{1}{r}{\textbf{64\%}} & \multicolumn{1}{r}{\textbf{78\%}} \\
        \midrule
            \multicolumn{7}{c}{\underline{Few-shot ($k=8$) settings}} \\
        \midrule
            \multirow{3}{*}{\textit{Amazon}} &
            \multicolumn{1}{c}{Vanilla} &
            \multicolumn{1}{r}{68\%} & \multicolumn{1}{r}{83\%} & \multicolumn{1}{r}{53\%} & \multicolumn{1}{r}{48\%} & \multicolumn{1}{r}{63\%} \\ \cline{2-7}
            & \multicolumn{1}{c}{CoT} &
            \multicolumn{1}{r}{70\%} & \multicolumn{1}{r}{85\%} & \multicolumn{1}{r}{55\%} & \multicolumn{1}{r}{64\%} & \multicolumn{1}{r}{68.50\%} \\ \cline{2-7}
            & \multicolumn{1}{c}{CoT$_\text{CR}$} &
            \multicolumn{1}{r}{\textbf{74\%}} & \multicolumn{1}{r}{\textbf{89\%}} & \multicolumn{1}{r}{\textbf{60\%}} & \multicolumn{1}{r}{\textbf{87\%}} & \multicolumn{1}{r}{\textbf{77.50\%}} \\ 
        \midrule
            \multirow{3}{*}{\textit{Google}} &
            \multicolumn{1}{c}{Vanilla} &
            \multicolumn{1}{r}{70\%} & \multicolumn{1}{r}{82\%} & \multicolumn{1}{r}{60\%} & \multicolumn{1}{r}{45\%} & \multicolumn{1}{r}{64.25\%} \\\cline{2-7}
            & \multicolumn{1}{c}{CoT} &
            \multicolumn{1}{r}{\textbf{71\%}} & \multicolumn{1}{r}{80\%} & \multicolumn{1}{r}{66\%} & \multicolumn{1}{r}{58\%} & \multicolumn{1}{r}{68.75\%} \\ \cline{2-7}
            & \multicolumn{1}{c}{CoT$_\text{CR}$} &
            \multicolumn{1}{r}{\textbf{71\%}} & \multicolumn{1}{r}{\textbf{84\%}} & \multicolumn{1}{r}{\textbf{70\%}} & \multicolumn{1}{r}{\textbf{78\%}} & \multicolumn{1}{r}{\textbf{75.75\%}} \\         
        \midrule
            \multirow{3}{*}{\textit{IMDB}} & 
            \multicolumn{1}{c}{Vanilla} &
            \multicolumn{1}{r}{67\%} & \multicolumn{1}{r}{58\%} & \multicolumn{1}{r}{85\%} & \multicolumn{1}{r}{28\%} & \multicolumn{1}{r}{59.50\%} \\ \cline{2-7}
            & \multicolumn{1}{c}{CoT} &
            \multicolumn{1}{r}{70\%} & \multicolumn{1}{r}{81\%} & \multicolumn{1}{r}{84\%} & \multicolumn{1}{r}{38\%} & \multicolumn{1}{r}{68.25\%} \\ \cline{2-7}
            & \multicolumn{1}{c}{CoT$_\text{CR}$} &
            \multicolumn{1}{r}{\textbf{71\%}} & \multicolumn{1}{r}{\textbf{88\%}} & \multicolumn{1}{r}{\textbf{89\%}} & \multicolumn{1}{r}{\textbf{97\%}} & \multicolumn{1}{r}{\textbf{86.25\%}} \\
        \bottomrule
    \end{tabular}
\end{table}

\section{Qualitative Evaluation}
In this section, we report three qualitative evaluations: an effectiveness test, a usability test, and expert feedback. All studies were conducted under an IRB-exempt protocol approved by Sejong University, Seoul, South Korea (SUIRB-HR-E-2024-016). Informed consent was obtained from all participants prior to participation.

\subsection{Effectiveness}
We recruited eight participants with backgrounds in data visualization: one undergraduate student ($PE_0$), one master's student ($PE_1$), two doctoral students ($PE_{2,3}$), and four participants holding master's degrees ($PE_{4\text{--}7}$). 
Each participant completed two tasks---manual coding and labeling using OPRA-Vis---on the same set of 20 questions (five per OPRA concept: \textit{Trust}, \textit{Satisfaction}, \textit{Commitment}, and \textit{Control Mutuality}) derived from the Amazon review dataset described in Section~\ref{subsec:case_amazon}. For both tasks, possible responses were \textit{``True''}, \textit{``False''}, or \textit{``I don't know''}.

\begin{table}[t]
    \footnotesize
    \caption{
        Comparison of OPRA labeling accuracy between human coding and our system. \textit{``I don't know''} counts appear in parentheses. Paired t-test shows $p<0.05$ (significant). {P.NO.} indicates participant ID.
    }
    \centering
    \label{table:acc_test_fully}
    \begin{tabular}{p{0.05\textwidth}p{0.125\textwidth}p{0.125\textwidth}p{0.125\textwidth}p{0.125\textwidth}p{0.05\textwidth}}
        \toprule
            \multicolumn{1}{c}{\multirow{2}{*}{\textbf{P.NO.}}} & \multicolumn{4}{c}{\textbf{OPRA Labeling Accuracy}} & \multicolumn{1}{c}{\textbf{Average}}\\ 
            & \multicolumn{1}{c}{\textit{Trust}} & \multicolumn{1}{c}{\textit{Satis.}} & \multicolumn{1}{c}{\textit{Commit.}} & \multicolumn{1}{c}{\textit{C.M.}} & \multicolumn{1}{c}{\textbf{Accuracy}} \\
        \midrule
            \multicolumn{6}{c}{\underline{Human-coding}} \\
        \midrule
            \multicolumn{1}{c}{$PE_0$} & \multicolumn{1}{r}{100\%} & \multicolumn{1}{r}{100\%} & \multicolumn{1}{r}{60\% (1)} & \multicolumn{1}{r}{100\%} & \multicolumn{1}{r}{90\% (1)} \\ 
            \multicolumn{1}{c}{$PE_4$} & \multicolumn{1}{r}{80\% (1)} & \multicolumn{1}{r}{100\%} & \multicolumn{1}{r}{40\% (2)} & \multicolumn{1}{r}{100\%} & \multicolumn{1}{r}{80\% (3)} \\ 
            \multicolumn{1}{c}{$PE_1$} & \multicolumn{1}{r}{100\%} & \multicolumn{1}{r}{100\%} & \multicolumn{1}{r}{40\% (2)} & \multicolumn{1}{r}{80\% (1)} & \multicolumn{1}{r}{80\% (3)} \\ 
            \multicolumn{1}{c}{$PE_6$} & \multicolumn{1}{r}{80\% (1)} & \multicolumn{1}{r}{100\%} & \multicolumn{1}{r}{20\% (3)} & \multicolumn{1}{r}{100\%} & \multicolumn{1}{r}{75\% (4)} \\ 
            \multicolumn{1}{c}{$PE_2$} & \multicolumn{1}{r}{80\% (1)} & \multicolumn{1}{r}{80\% (1)} & \multicolumn{1}{r}{40\% (3)} & \multicolumn{1}{r}{80\% (1)} & \multicolumn{1}{r}{70\% (6)} \\ 
            \multicolumn{1}{c}{$PE_3$} & \multicolumn{1}{r}{60\% (1)} & \multicolumn{1}{r}{80\% (1)} & \multicolumn{1}{r}{80\%} & \multicolumn{1}{r}{100\%} & \multicolumn{1}{r}{80\% (2)} \\ 
            \multicolumn{1}{c}{$PE_5$} & \multicolumn{1}{r}{100\%} & \multicolumn{1}{r}{100\%} & \multicolumn{1}{r}{60\% (1)} & \multicolumn{1}{r}{100\%} & \multicolumn{1}{r}{90\% (1)} \\ 
            \multicolumn{1}{c}{$PE_7$} & \multicolumn{1}{r}{80\% (1)} & \multicolumn{1}{r}{80\%} & \multicolumn{1}{r}{60\% (2)} & \multicolumn{1}{r}{100\%} & \multicolumn{1}{r}{80\% (3)} \\ 
            \multicolumn{1}{c}{Average} & \multicolumn{1}{r}{85\%} & \multicolumn{1}{r}{92.5\%} & \multicolumn{1}{r}{50\%} & \multicolumn{1}{r}{\textbf{95\%}} & \multicolumn{1}{r}{80.63\%} \\
        \midrule
            \multicolumn{6}{c}{\underline{Using our system}} \\
        \midrule
            \multicolumn{1}{c}{$PE_0$} & \multicolumn{1}{r}{100\%} & \multicolumn{1}{r}{100\%} & \multicolumn{1}{r}{100\%} & \multicolumn{1}{r}{80\%} & \multicolumn{1}{r}{95\%} \\ 
            \multicolumn{1}{c}{$PE_1$} & \multicolumn{1}{r}{100\%} & \multicolumn{1}{r}{100\%} & \multicolumn{1}{r}{80\%} & \multicolumn{1}{r}{80\%} & \multicolumn{1}{r}{90\%} \\ 
            \multicolumn{1}{c}{$PE_4$} & \multicolumn{1}{r}{100\%} & \multicolumn{1}{r}{100\%} & \multicolumn{1}{r}{100\%} & \multicolumn{1}{r}{100\%} & \multicolumn{1}{r}{100\%} \\ 
            \multicolumn{1}{c}{$PE_6$} & \multicolumn{1}{r}{80\%} & \multicolumn{1}{r}{100\%} & \multicolumn{1}{r}{100\%} & \multicolumn{1}{r}{80\%} & \multicolumn{1}{r}{90\%} \\ 
            \multicolumn{1}{c}{$PE_2$} & \multicolumn{1}{r}{80\%} & \multicolumn{1}{r}{100\%} & \multicolumn{1}{r}{80\%} & \multicolumn{1}{r}{80\%} & \multicolumn{1}{r}{85\%} \\ 
            \multicolumn{1}{c}{$PE_3$} & \multicolumn{1}{r}{100\%} & \multicolumn{1}{r}{100\%} & \multicolumn{1}{r}{80\% (1)} & \multicolumn{1}{r}{100\%} & \multicolumn{1}{r}{95\% (1)} \\ 
            \multicolumn{1}{c}{$PE_5$} & \multicolumn{1}{r}{100\%} & \multicolumn{1}{r}{100\%} & \multicolumn{1}{r}{100\%} & \multicolumn{1}{r}{80\%} & \multicolumn{1}{r}{95\%} \\ 
            \multicolumn{1}{c}{$PE_7$} & \multicolumn{1}{r}{100\%} & \multicolumn{1}{r}{80\%} & \multicolumn{1}{r}{80\%} & \multicolumn{1}{r}{100\%} & \multicolumn{1}{r}{90\%} \\ 
            \multicolumn{1}{c}{Average} & \multicolumn{1}{r}{\textbf{95\%}} & \multicolumn{1}{r}{\textbf{97.5\%}} & \multicolumn{1}{r}{\textbf{90\%}} & \multicolumn{1}{r}{87.5\%} & \multicolumn{1}{r}{\textbf{92.5}} \\
        \bottomrule
    \end{tabular}
\end{table}

Table~\ref{table:acc_test_fully} summarizes labeling accuracy. OPRA-Vis outperformed human coding for \textit{Trust}, \textit{Satisfaction}, and \textit{Commitment}, with a 40\% gain in \textit{Commitment}. It also reduced \textit{``I don't know''} responses by 82.61\%. A paired t-test confirmed significance ($p = 0.0002 < 0.05$). However, accuracy for \textit{Control Mutuality} decreased by 7.5\% compared to human coding.

The accuracy drop was due to \textit{Control Mutuality} being assessed in a review (\textit{``I honestly haven't sought out much information about the governance of sprint. I'm sure they are willing to be transparent, I'm just not all that interested''}). Participants initially labeled it \textit{False} but changed to \textit{True} after viewing OPRA-Vis's reasoning. PR experts clarified that interpretation depends on whether emphasis is placed on the first or second sentence in the reviews. The experts prioritized the first sentence, reasoning that insufficient customer knowledge of company efforts undermines any basis for assessing \textit{Control Mutuality}. They noted, \textit{``Uncertainty about information alone does not confirm adequate organizational effort to maintain control mutuality''}. However, they acknowledged that emphasizing the second sentence could make OPRA-Vis's decision appear reasonable. Ultimately, given linguistic nuances in PR contexts, they classified \textit{Control Mutuality} as \textit{False}. This discrepancy highlights a limitation of OPRA-Vis's in-context learning, discussed in Section~\ref{sec:discussion}.

\subsection{Usability}
We recruited 12 participants with diverse domain knowledge to evaluate OPRA-Vis usability: Mathematics, computer science, and PR ($PU_0$); computer science and PR ($PU_1$); computer science and NLP ($PU_{2,3}$); game development ($PU_4$); traffic and deep learning ($PU_5$); and computer science ($PU_{6\textrm{--}11}$). All were familiar with online reviews and experienced with data analysis tools (e.g., Excel, Tableau, IBM SPSS) and Python tool development. $PU_0$ and $PU_1$ had PR-related development experience; $PU_0$, $PU_1$, $PU_5$, and $PU_6$ had data visualization knowledge. None participated in developing OPRA-Vis. The usability test used the same dataset as in Section~\ref{sec:quantitative_evaluation}.

The usability test had three stages: description, utilization, and feedback. Participants first learned about PR theory, CoT prompting, and public opinion data. Then, they used OPRA-Vis for 45 minutes to explore scenarios from Section~\ref{sec:case_study}, adjusting labels via prompt editing. Finally, they completed a post-test system usability scale (SUS) questionnaire~\cite{brooke1996sus} and provided feedback. In the usability test, OPRA-Vis received an average SUS score of 84.75, with a median of 90, a minimum of 67.5, and a maximum of 95. Given that scores above 70 are generally considered acceptable~\cite{bangor2008an}, these results indicate that OPRA-Vis is perceived as highly usable. For each item of the SUS questionnaire ($I_{1\textrm{--}10}$), participants gave 90 ($I_1$), 90 ($I_2$), 90 ($I_3$), 67.5 ($I_4$), 90 ($I_5$), 77.5 ($I_6$), 90 ($I_7$), 85 ($I_8$), 95 ($I_9$), and 72.5 ($I_{10}$). Among them, $I_4$ (\textit{``I think that I would need the support of a technical person to be able to use this system''}), $I_6$ (\textit{``I thought there was too much inconsistency in this system''}), and $I_{10}$ (\textit{``I needed to learn a lot of things before I could get going with this system''}) were evaluated as having low contribution compared to other items.

We identified reasons for low contributions on three items based on participant feedback. For item $I_4$, participants ($PU_{2,4,7}$) struggled due to limited understanding of data visualization. For $I_6$, participants ($PU_{4,7,11}$) noted inconsistencies in LLM decisions versus ground-truth, attributed to low comprehension of PR theory. For $I_{10}$, participants ($PU_{0,2,8}$) raised usability concerns: $PU_0$ highlighted challenges for non-PR users, $PU_2$ noted difficulties in writing prompts for NLP novices, and $PU_8$ cited limited knowledge in various areas. They suggested that a simplified OPRA-Vis could improve usability for non-experts. Positive feedback included comments such as, \textit{``It is a tool [...] to easily learn and utilize PR knowledge''}, and improvement suggestions included, \textit{``If there were a system that showed the difference between pre-feedback and post-feedback, [...] feedback would be clearer''}.

\subsection{Expert Feedback}
\label{sec:expert_feedback}
Seven experts evaluated OPRA-Vis's effectiveness and suggested improvements. Two specialize in PR theory ($E_0$, $E_1$), one in PR-based deep learning model development ($E_2$), one in PR-based visual analysis ($E_3$), two in NLP ($E_4$, $E_5$), and one in data visualization ($E_6$). $E_1$ and $E_2$ contributed to OPRA-Vis's design and ground truth generation; others encountered OPRA-Vis for the first time after development.

\textbf{Effectiveness:}
Experts praised the intuitive visualizations. In tag clouds, common words or company names such as \textit{``type, five, trip, Amazon, Apple''} are typically excluded from sentiment analysis because they lack inherent semantic meaning. However, in public opinion data, these words can carry significance. For example, in the tag cloud for \textit{``Trust=True''} shown in Figure~\ref{fig:case_amazon}~(a), \textit{``Amazon''} exhibits positive sentiment, as it frequently appears in contexts like \textit{``I order a lot of stuff from Amazon.
They always deliver on time''}, which provides \textit{True} clues for \textit{Trust}. $E_0$ noted, \textit{``The bipolar illustration of key relational dimensions like commitment and trust allows PR managers to perceive and articulate the strength and tendencies in data distributions of relationships''}. $E_6$ remarked, \textit{``This view aids in understanding overall opinion distributions and common keywords. It serves as a starting point for selecting relevant comments to verify if the LLM operated as intended''}. Both $E_3$ and $E_6$ highlighted the system's effectiveness: \textit{``This view promptly helps understand how LLM evaluates each comment. The displayed texts of clues and reasoning are useful for understanding how LLM labeled the comment. [...]; In particular [...], attention visualization greatly assists reasoning''}. Moreover, $E_1$ emphasized, \textit{``[...] it is important to note that human coding can often be inconsistent and not always accurate''}, and concluded that OPRA-Vis successfully addresses the limitations of existing human coding.

\textbf{Improvements:}
Experts suggested extending OPRA-Vis to include time-series analysis of public opinion, beyond its current focus on reasoning and correction. $E_0$ noted, \textit{``Time-series data is also critical. [...] Real-time tracking enables managers to observe historical changes in relationships. [...] can inform PR managers about events triggering changes (\textit{True} or \textit{False}) and allow for the preparation of appropriate managerial and communication responses''}. $E_5$ recommended adding automation and visual reliability cues, stating, \textit{``Wouldn't it be possible to approach things in a more user-friendly way by leaving the variables untouchable and, if there is nothing to use, pressing None, as LLM solves the problem? [...] I think that if the reliability of the LLM is drawn in a score or graph, we can rely on the LLM and use it''}. Future work will focus on generating labeled data from OPRA-Vis to enable fine-tuning and exploring visual analysis techniques for tracking opinion trends over time.
\section{Conclusion and Discussion}
\label{sec:discussion}
This paper introduced OPRA-Vis, a visual analytics system supporting organization-public relationship assessment~(OPRA) by leveraging LLMs and CoT prompting for expert-driven labeling. Integrating PR's OPRA concepts, it semi-automates public opinion labeling with CoT prompting enriched by expert-informed clues and reasoning. OPRA-Vis addresses the linguistic challenges and limitations of CoT prompting by visualizing LLM's reasoning for labeling using the certainty of concepts ($CoC$), sentiment, and attention scores. The system supports interactive analysis using the OPRA Concept Space view, Data Management view, and LLM Reasoning \& Editing view. Evaluations, which include use scenarios, quantitative tests, usability studies, and expert feedback, demonstrate an effective integration of expert knowledge with LLM reasoning without the need for fine-tuning, although some limitations remain.

\textbf{In-context learning:} In-context learning through CoT prompting with \textit{few-shot examples (FSEs)} is widely used and can be useful for reasoning tasks~\cite{brown2020language}, but often underperforms fine-tuned models in text assessment due to language complexity~\cite{sun2023text}. Its effectiveness heavily depends on the structure and content of prompts, requiring domain expertise~\cite{wang2021want}. However, domain experts often need assistance with LLM-specific techniques. To address linguistic challenges and improve reliability, we integrated visual analysis to clarify LLM reasoning. We incorporated expert knowledge by defining OPRA concepts, including ten definitions of top-level and sub-dimensions~\cite{grunig2001guidelines}. To ensure clarity and relevance, we adopted few-shot CoT prompting with clues and reasons~\cite{sun2023text}. Still, further work is needed to improve reasoning in in-context learning, and to address the constraints of current fine-tuning techniques, e.g., parameter-efficient tuning~\cite{ding2023parameter}.

\begin{figure}[t]
    \centering
    \includegraphics[width=\linewidth]{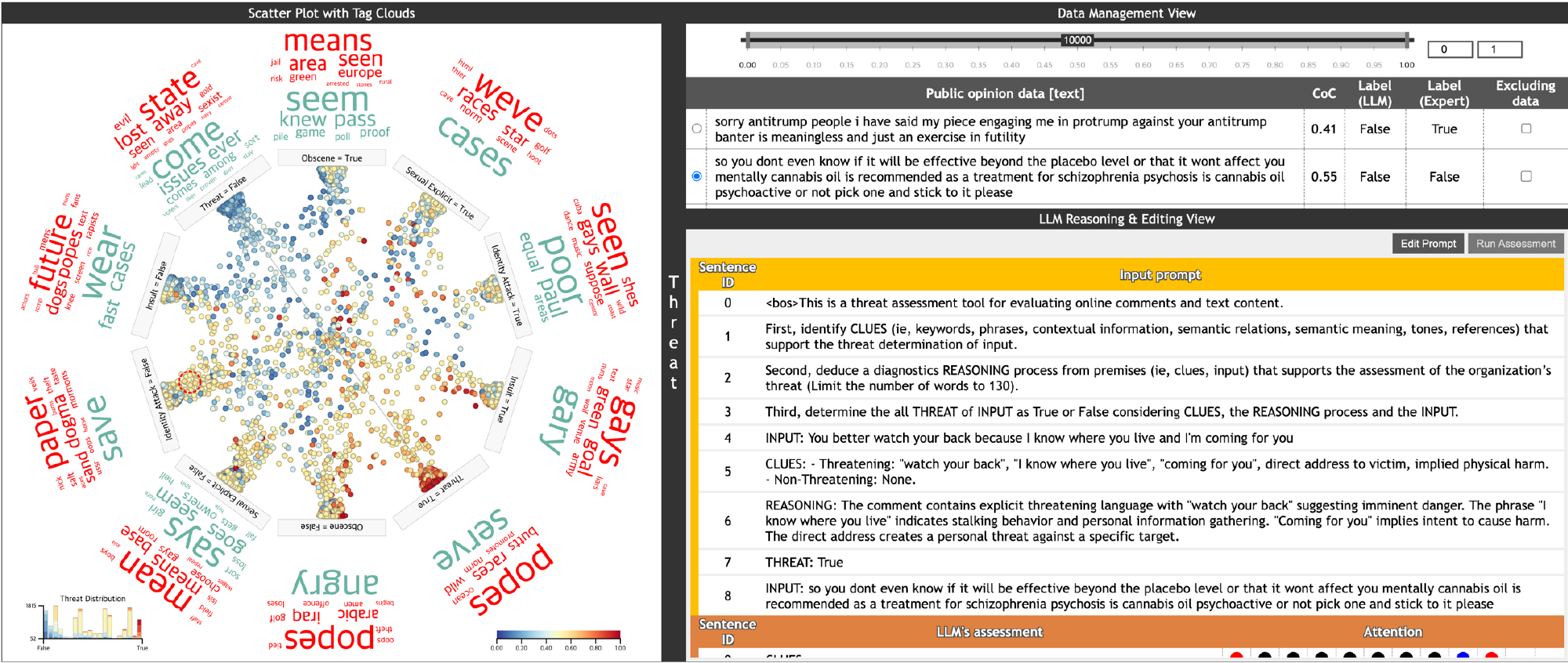}
    \caption{%
        Application of OPRA-Vis to Jigsaw toxic comments dataset.
    }
    \label{fig:dis_domain}
\end{figure}

\textbf{Domain applicability:} While OPRA-Vis is designed to address challenges in the PR domain, its framework can be adapted to other domains. As shown in Figure~\ref{fig:dis_domain}, we applied OPRA-Vis to the Jigsaw toxic comments dataset~\cite{kaggle2019jigsaw}, which has five dimensions. Datasets with more than five dimensions challenge the scatter plot with tag clouds, as \textit{True} and \textit{False} labels are arranged symmetrically ($5 \times 2$ sides). With more sides, the layout becomes circular, making it harder to interpret relationships. Future work will develop a visualization to handle multiple dimensions effectively.

\section*{Acknowledgment}
\noindent This should be a simple paragraph before the bibliography to thank those individuals and institutions who have supported your work on this article.

\bibliographystyle{IEEEtran}

\bibliography{0_TVCG_PR_LLM}

\begin{thebibliography}{10}
\providecommand{\url}[1]{#1}
\csname url@samestyle\endcsname
\providecommand{\newblock}{\relax}
\providecommand{\bibinfo}[2]{#2}
\providecommand{\BIBentrySTDinterwordspacing}{\spaceskip=0pt\relax}
\providecommand{\BIBentryALTinterwordstretchfactor}{4}
\providecommand{\BIBentryALTinterwordspacing}{\spaceskip=\fontdimen2\font plus
\BIBentryALTinterwordstretchfactor\fontdimen3\font minus \fontdimen4\font\relax}
\providecommand{\BIBforeignlanguage}[2]{{%
\expandafter\ifx\csname l@#1\endcsname\relax
\typeout{** WARNING: IEEEtran.bst: No hyphenation pattern has been}%
\typeout{** loaded for the language `#1'. Using the pattern for}%
\typeout{** the default language instead.}%
\else
\language=\csname l@#1\endcsname
\fi
#2}}
\providecommand{\BIBdecl}{\relax}
\BIBdecl

\bibitem{kim2011problem}
J.-N. Kim and J.~E. Grunig, ``Problem solving and communicative action: A situational theory of problem solving,'' \emph{Journal of Communication}, vol.~61, no.~1, pp. 120--149, 2011.

\bibitem{kim2013strategic}
J.-N. Kim, C.-J.~F. Hung-Baesecke, S.-U. Yang, and J.~E. Grunig, ``A strategic management approach to reputation, relationships, and publics: the research heritage of the excellence theory$^1$,'' \emph{The Handbook of Communication and Corporate Reputation}, pp. 197--212, 2013.

\bibitem{grunig2017publics}
J.~E. Grunig and J.-N. Kim, ``Publics approaches to segmentation in health and risk messaging,'' in \emph{The Oxford Encyclopedia of Health and Risk Message Design and Processing}, R.~L. Parrott, Ed.\hskip 1em plus 0.5em minus 0.4em\relax Oxford University Press, 2017, (37 pages).

\bibitem{kim2014lay}
J.-N. Kim, ``Lay consumer informatics and fast-choicism: Instant trust and prompt loyalty in digitalized, networked marketplace,'' \emph{Communication Insight}, vol.~3, pp. 10--31, 2014.

\bibitem{kim2021pseudo}
J.-N. Kim and H.~G. de~Z{\'u}{\~n}iga, ``Pseudo-information, media, publics, and the failing marketplace of ideas: Theory,'' \emph{American Behavioral Scientist}, vol.~65, no.~2, pp. 163--179, 2021.

\bibitem{grunig2002excellent}
J.~E. Grunig and D.~M. Dozier, \emph{Excellent Public Relations and Effective Organizations: A Study of Communication Management in Three Countries}.\hskip 1em plus 0.5em minus 0.4em\relax Routledge, 2002.

\bibitem{hon1999guidelines}
L.~C. Hon and J.~E. Grunig, ``Guidelines for measuring relationships in public relations,'' Institute for Public Relations, 1999.

\bibitem{huang2001opra}
Y.-H. Huang, ``{OPRA}: A cross-cultural, multiple-item scale for measuring organization-public relationships,'' \emph{Journal of Public Relations Research}, vol.~13, no.~1, pp. 61--90, 2001.

\bibitem{kim2013integrating}
J.-N. Kim and L.~Ni, ``Integrating formative and evaluative research in two types of public relations problems: A review of research programs within the behavioral, strategic management paradigm,'' \emph{Journal of Public Relations Research}, vol.~25, no.~1, pp. 1--29, 2013.

\bibitem{beltagy2019scibert}
I.~Beltagy, K.~Lo, and A.~Cohan, ``{S}ci{BERT}: A pretrained language model for scientific text,'' in \emph{Proc. EMNLP-IJCNLP}.\hskip 1em plus 0.5em minus 0.4em\relax ACL, 2019, pp. 3615--3620.

\bibitem{gu2022domain}
Y.~Gu, R.~Tinn, H.~Cheng, M.~Lucas, N.~Usuyama \emph{et~al.}, ``Domain-specific language model pretraining for biomedical natural language processing,'' \emph{ACM Trans. Comput. Healthcare}, vol.~3, no.~1, oct 2021.

\bibitem{howard2018universal}
J.~Howard and S.~Ruder, ``Universal language model fine-tuning for text classification,'' in \emph{Proc. ACL}.\hskip 1em plus 0.5em minus 0.4em\relax ACL, 2018, pp. 328--339.

\bibitem{grunig2001guidelines}
J.~E. Grunig and L.~A. Grunig, ``Guidelines for formative and evaluative research in public affairs,'' \emph{A Report for the Department of Energy Office of Science}, 2001.

\bibitem{wang2021want}
S.~Wang, Y.~Liu, Y.~Xu, C.~Zhu, and M.~Zeng, ``Want to reduce labeling cost? {GPT}-3 can help,'' in \emph{Proc. Findings of EMNLP}.\hskip 1em plus 0.5em minus 0.4em\relax ACL, 2021, pp. 4195--4205.

\bibitem{sun2023text}
X.~Sun, X.~Li, J.~Li, F.~Wu, S.~Guo \emph{et~al.}, ``Text classification via large language models,'' in \emph{Proc. Findings of EMNLP}.\hskip 1em plus 0.5em minus 0.4em\relax ACL, 2023, pp. 8990--9005.

\bibitem{zhao2024explainability}
H.~Zhao, H.~Chen, F.~Yang, N.~Liu, H.~Deng \emph{et~al.}, ``Explainability for large language models: A survey,'' \emph{ACM Trans Intell Syst Technol}, vol.~15, no.~2, 2024.

\bibitem{team2024gemma}
{Gemma Team}, ``Gemma: Open models based on {Gemini} research and technology,'' \emph{arXiv:2403.08295}, 2024.

\bibitem{brown2020language}
T.~Brown, B.~Mann, N.~Ryder, M.~Subbiah, J.~D. Kaplan \emph{et~al.}, ``Language models are few-shot learners,'' in \emph{Proc. NeurIPS}, vol.~33, 2020, pp. 1877--1901.

\bibitem{devlin2018bert}
J.~Devlin, M.-W. Chang, K.~Lee, and K.~Toutanova, ``{BERT}: Pre-training of deep bidirectional transformers for language understanding,'' in \emph{Proc. NAACL}.\hskip 1em plus 0.5em minus 0.4em\relax ACL, 2019, pp. 4171--4186.

\bibitem{achiam2023gpt}
J.~Achiam, S.~Adler, S.~Agarwal, L.~Ahmad, I.~Akkaya \emph{et~al.}, ``{GPT-4} technical report,'' \emph{arXiv:2303.08774}, 2023.

\bibitem{dong2022survey}
Q.~Dong, L.~Li, D.~Dai, C.~Zheng, Z.~Wu \emph{et~al.}, ``A survey on in-context learning,'' \emph{arXiv:2301.00234}, 2022.

\bibitem{chen2022improving}
M.~Chen, J.~Du, R.~Pasunuru, T.~Mihaylov, S.~Iyer \emph{et~al.}, ``Improving in-context few-shot learning via self-supervised training,'' in \emph{Proc. NAACL}.\hskip 1em plus 0.5em minus 0.4em\relax ACL, 2022, pp. 3558--3573.

\bibitem{wei2022chain}
J.~Wei, X.~Wang, D.~Schuurmans, M.~Bosma, F.~Xia \emph{et~al.}, ``Chain-of-thought prompting elicits reasoning in large language models,'' \emph{Proc. NeurIPS}, vol.~35, pp. 24\,824--24\,837, 2022.

\bibitem{wu2023chain}
D.~Wu, J.~Zhang, and X.~Huang, ``Chain of thought prompting elicits knowledge augmentation,'' in \emph{Proc. Findings of EMNLP}.\hskip 1em plus 0.5em minus 0.4em\relax ACL, 2023, pp. 6519--6534.

\bibitem{perez2021true}
E.~Perez, D.~Kiela, and K.~Cho, ``True few-shot learning with language models,'' \emph{Proc. NeurIPS}, vol.~34, pp. 11\,054--11\,070, 2021.

\bibitem{ammar2018construction}
W.~Ammar, D.~Groeneveld, C.~Bhagavatula, I.~Beltagy, M.~Crawford \emph{et~al.}, ``Construction of the literature graph in semantic scholar,'' in \emph{Proc. NAACL}.\hskip 1em plus 0.5em minus 0.4em\relax ACL, 2018, pp. 84--91.

\bibitem{meng2020text}
Y.~Meng, Y.~Zhang, J.~Huang, C.~Xiong, H.~Ji \emph{et~al.}, ``Text classification using label names only: A language model self-training approach,'' in \emph{Proc. EMNLP}.\hskip 1em plus 0.5em minus 0.4em\relax ACL, 2020, pp. 9006--9017.

\bibitem{keim2008visual}
D.~Keim, G.~Andrienko, J.-D. Fekete, C.~G{\"o}rg, J.~Kohlhammer, and G.~Melan{\c{c}}on, ``Visual analytics: Definition, process, and challenges,'' in \emph{Information Visualization: Human-Centered Issues and Perspectives}, A.~Kerren, J.~T. Stasko, J.-D. Fekete, and C.~North, Eds.\hskip 1em plus 0.5em minus 0.4em\relax Springer, 2008, pp. 154--175.

\bibitem{liu2019bridging}
S.~Liu, X.~Wang, C.~Collins, W.~Dou, F.~Ouyang \emph{et~al.}, ``Bridging text visualization and mining: A task-driven survey,'' \emph{IEEE Trans Vis and Comput Graph}, vol.~25, no.~7, pp. 2482--2504, 2019.

\bibitem{knittel2021pyramidtags}
J.~Knittel, S.~Koch, and T.~Ertl, ``{PyramidTags}: Context-, time- and word order-aware tag maps to explore large document collections,'' \emph{IEEE Trans. Vis. Comput. Graph.}, vol.~27, no.~12, pp. 4455--4468, 2021.

\bibitem{khayat2020vassl}
M.~Khayat, M.~Karimzadeh, J.~Zhao, and D.~S. Ebert, ``{VASSL}: A visual analytics toolkit for social spambot labeling,'' \emph{IEEE Trans. Vis. Comput. Graph.}, vol.~26, no.~1, pp. 874--883, 2020.

\bibitem{li2024evovis}
S.~Li, G.~Liu, T.~Wei, S.~Jia, and J.~Zhang, ``{EvoVis}: A visual analytics method to understand the labeling iterations in data programming,'' \emph{IEEE Trans. Vis. Comput. Graph.}, pp. 1--16, 2024 (Early Access).

\bibitem{vig2019multiscale}
J.~Vig, ``A multiscale visualization of attention in the transformer model,'' in \emph{Proc. ACL}.\hskip 1em plus 0.5em minus 0.4em\relax ACL, 2019, pp. 37--42.

\bibitem{yeh2024attentionviz}
C.~Yeh, Y.~Chen, A.~Wu, C.~Chen, F.~Viégas, and M.~Wattenberg, ``{AttentionViz}: A global view of transformer attention,'' \emph{IEEE Trans. Vis. Comput. Graph.}, vol.~30, no.~1, pp. 262--272, 2024.

\bibitem{coscia2023knowledgevis}
A.~Coscia and A.~Endert, ``{KnowledgeVIS}: Interpreting language models by comparing fill-in-the-blank prompts,'' \emph{IEEE Trans. Vis. Comput. Graph.}, pp. 1--13, 2023 (Early Access).

\bibitem{guo2024prompthis}
Y.~Guo, H.~Shao, C.~Liu, K.~Xu, and X.~Yuan, ``Prompthis: Visualizing the process and influence of prompt editing during text-to-image creation,'' \emph{IEEE Trans. Vis. Comput. Graph.}, pp. 1--12, 2024.

\bibitem{brandwatch}
Brandwatch, ``Brandwatch. monitor \& analyze the mentions of your brand with social media listening tool.'' \url{https://www.brandwatch.com/company/about/}, [Last visited: November 25, 2024].

\bibitem{hayes2021can}
J.~L. Hayes, B.~C. Britt, W.~Evans, S.~W. Rush, N.~A. Towery, and A.~C. Adamson, ``Can social media listening platforms’ artificial intelligence be trusted? examining the accuracy of crimson hexagon’s (now brandwatch consumer research’s) ai-driven analyses,'' \emph{Journal of Advertising}, vol.~50, no.~1, pp. 81--91, 2021.

\bibitem{mohamed2022analyzing}
K.~Mohamed and {\"U}.~A. Bayraktar, ``Analyzing the role of sentiment analysis in public relations: Brand monitoring and crisis management,'' \emph{International Journal of Humanities and Social Science}, vol.~9, no.~3, pp. 116--126, 2022.

\bibitem{himelboim2014asocial}
I.~Himelboim, G.~J. Golan, B.~B. Moon, and R.~J. Suto, ``A social networks approach to public relations on {Twitter}: Social mediators and mediated public relations,'' \emph{Journal of Public Relations Research}, vol.~26, no.~4, pp. 359--379, 2014.

\bibitem{seltzer2010toward}
T.~Seltzer and W.~Zhang, ``Toward a model of political organization--public relationships: Antecedent and cultivation strategy influence on citizens' relationships with political parties,'' \emph{Journal of Public Relations Research}, vol.~23, no.~1, pp. 24--45, 2010.

\bibitem{lee2013explicating}
H.~M. Lee and J.~W. Jun, ``Explicating public diplomacy as organization--public relationship ({OPR}): An empirical investigation of {OPR}s between the {US} embassy in {Seoul} and {South Korean} college students,'' \emph{Journal of Public Relations Research}, vol.~25, no.~5, pp. 411--425, 2013.

\bibitem{chon2021predicting}
M.-G. Chon and H.~Park, ``Predicting public support for government actions in a public health crisis: Testing fear, organization-public relationship, and behavioral intention in the framework of the situational theory of problem solving,'' \emph{Health Communication}, vol.~36, no.~4, pp. 476--486, 2021.

\bibitem{bortree2010exploring}
D.~S. Bortree, ``Exploring adolescent--organization relationships: A study of effective relationship strategies with adolescent volunteers,'' \emph{Journal of Public Relations Research}, vol.~22, no.~1, pp. 1--25, 2010.

\bibitem{mccorkindale2010can}
T.~McCorkindale, ``Can you see the writing on my wall? {A} content analysis of the {Fortune 50}’s {Facebook} social networking sites,'' \emph{Public Relations Journal}, vol.~4, no.~3, pp. 1--13, 2010.

\bibitem{linvill2012colleages}
D.~L. Linvill, S.~E. McGee, and L.~K. Hicks, ``Colleges’ and universities’ use of {Twitter}: A content analysis,'' \emph{Public Relations Review}, vol.~38, no.~4, pp. 636--638, 2012.

\bibitem{andohquainoo2015theuseof}
L.~Andoh-Quainoo and P.~Annor-Antwi, ``The use of social media in public relations: A case of {Facebook} in the {Ghanaian} financial services industry,'' \emph{New Media and Mass Communication}, vol.~41, pp. 37--47, 2015.

\bibitem{tsne2008}
L.~van~der Maaten and G.~Hinton, ``Visualizing data using {t-SNE},'' \emph{J Mach Learn Res}, vol.~9, no.~86, pp. 2579--2605, 2008.

\bibitem{newton1833principia}
I.~Newton, \emph{Philosophiae Naturalis Principia Mathematica}.\hskip 1em plus 0.5em minus 0.4em\relax G. Brookman, 1833.

\bibitem{seo2025sentence}
S.~Seo, S.~Yoo, H.~Lee, Y.~Jang, J.~H. Park, and J.-N. Kim, ``A sentence-level visualization of attention in large language models,'' in \emph{Proc. NAACL}.\hskip 1em plus 0.5em minus 0.4em\relax ACL, 2025, pp. 313--320.

\bibitem{amazon}
I.~o. i.~a. Amazon.com, ``Amazon.com. spend less. smile more,'' \url{https://www.amazon.com/ref=nav_logo}, 1996, 1996-2024.

\bibitem{li2022uctopic}
J.~Li, J.~Shang, and J.~McAuley, ``{UCT}opic: Unsupervised contrastive learning for phrase representations and topic mining,'' in \emph{Proc. ACL}.\hskip 1em plus 0.5em minus 0.4em\relax ACL, 2022, pp. 6159--6169.

\bibitem{garcia2023dataset}
I.~Garc{\'i}a-Ferrero, B.~Altuna, J.~Alvez, I.~Gonzalez-Dios, and G.~Rigau, ``This is not a dataset: A large negation benchmark to challenge large language models,'' in \emph{Proc. EMNLP}.\hskip 1em plus 0.5em minus 0.4em\relax ACL, 2023, pp. 8596--8615.

\bibitem{maas2011learning}
A.~L. Maas, R.~E. Daly, P.~T. Pham, D.~Huang, A.~Y. Ng, and C.~Potts, ``Learning word vectors for sentiment analysis,'' in \emph{Proc. ACL}.\hskip 1em plus 0.5em minus 0.4em\relax ACL, 2011, pp. 142--150.

\bibitem{touvron2023llama}
H.~Touvron, L.~Martin, K.~Stone, P.~Albert, A.~Almahairi \emph{et~al.}, ``Llama 2: Open foundation and fine-tuned chat models,'' \emph{arXiv:2307.09288}, 2023.

\bibitem{brooke1996sus}
J.~Brooke, ``{SUS}: A ``quick and dirty'' usability scale,'' in \emph{Usability Evaluation in Industry}, P.~W. Jordan, B.~Thomas, I.~L. McClelland, and B.~Weerdmeester, Eds.\hskip 1em plus 0.5em minus 0.4em\relax CRC Press, 1996, pp. 189--194.

\bibitem{bangor2008an}
P.~T.~K. Aaron~Bangor and J.~T. Miller, ``An empirical evaluation of the system usability scale,'' \emph{International Journal of Human–Computer Interaction}, vol.~24, no.~6, pp. 574--594, 2008.

\bibitem{ding2023parameter}
N.~Ding, Y.~Qin, G.~Yang, F.~Wei, Z.~Yang \emph{et~al.}, ``Parameter-efficient fine-tuning of large-scale pre-trained language models,'' \emph{Nature Machine Intelligence}, vol.~5, no.~3, pp. 220--235, 2023.

\bibitem{kaggle2019jigsaw}
Kaggle, ``Jigsaw unintended bias in toxicity classification,'' \url{https://kaggle.com/c/jigsaw-unintended-bias-in-toxicity-classification}, [Last visited: February 28, 2025].

\end{thebibliography}

\begin{IEEEbiography}[{\includegraphics[width=1in,height=1.25in,clip,keepaspectratio]{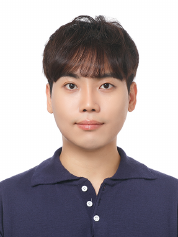}}]{Sangbong Yoo}
    received the bachelor's and Ph.D. degree in computer engineering from Sejong University, Seoul, South Korea, in 2015 and 2022, respectively. He was a postdoctoral researcher at Sejong University, from 2022 to 2025. He is currently a postdoctoral researcher at Korea Institute of Science and Technology (KIST), Seoul, South Korea. His research interests include information visualization, visual analytics, eye-gaze analysis, data quality, and volume rendering.
\end{IEEEbiography}

\begin{IEEEbiography}[{\includegraphics[width=1in,height=1.25in,clip,keepaspectratio]{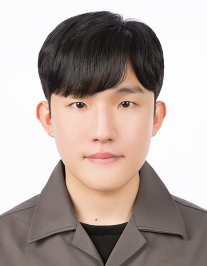}}]{Seongbum Seo}
    is a researcher at the Data Visualization Lab, Sejong University, Seoul, South Korea. He received the B.S. and M.S. degrees in computer engineering from Sejong University, in 2023 and 2025, respectively. His research interests include natural language processing and data visualization.
\end{IEEEbiography}

\begin{IEEEbiography}[{\includegraphics[width=1in,height=1.25in,clip,keepaspectratio]{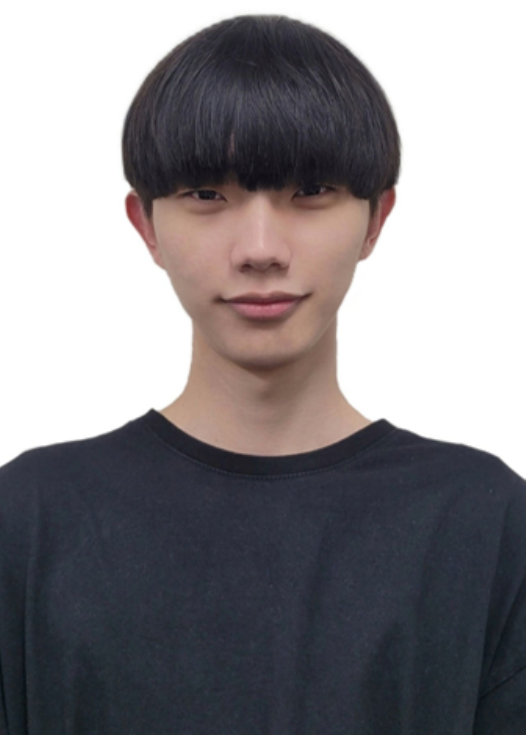}}]{Chanyoung Yoon}
    received a bachelor's and master's degree in computer engineering from Sejong University, Seoul, South Korea, in 2022 and 2023, respectively. He is now pursuing a Ph.D. degree at Sejong University, focusing on reinforcement learning and visual analytics. His research aims to deep reinforcement learning, visual analytics, and traffic prediction.
\end{IEEEbiography}

\begin{IEEEbiography}[{\includegraphics[width=1in,height=1.25in,clip,keepaspectratio]{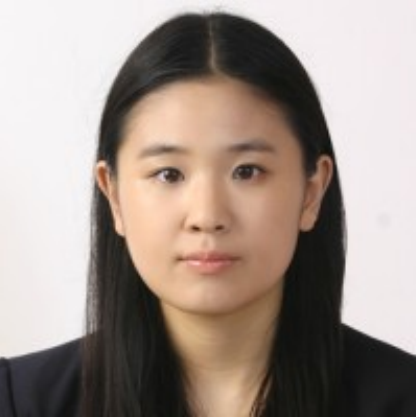}}]{Hyelim Lee}
     is an Assistant Professor at the School of Media and Communication at Korea University. She is also a senior researcher at the Debiasing and Lay Informatics (DaLI) Lab, focusing on public relations, public policy-making processes, and computational social science research projects. Previously, she worked at the Edward R. Murrow College of Communication at Washington State University.
\end{IEEEbiography}

\begin{IEEEbiography}[{\includegraphics[width=1in,height=1.25in,clip,keepaspectratio]{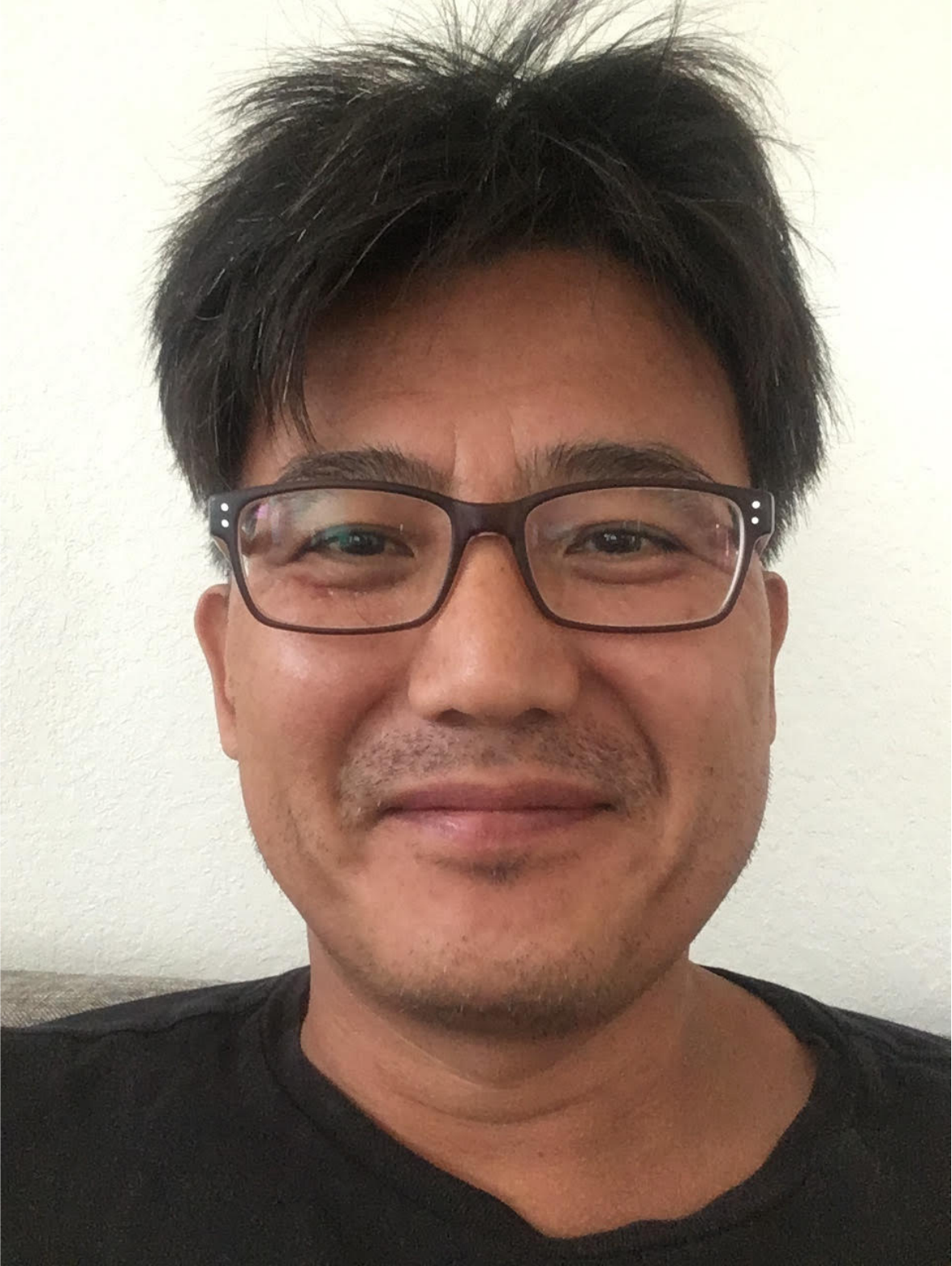}}]{Jeong-Nam Kim}
    (PhD, University of Maryland, College Park) is a communication theorist known for developing the Situational Theory of Problem Solving (STOPS) and for the model of cognitive arrest and epistemic inertia. He directs the DaLI (Debiasing \& Lay Informatics) Lab, which addresses challenges such as pseudo‐information, public bias, data and algorithmic bias, and dysfunctional information markets. Kim is a KAIST Chair Professor at the Moon Soul Graduate School of Future Strategy and the Kim Jaechul Graduate School of AI at KAIST. He previously held the Gaylord Family Endowed Chair at the University of Oklahoma and is a fellow of several international research centers.
\end{IEEEbiography}

\begin{IEEEbiography}[{\includegraphics[width=1in,height=1.25in,clip,keepaspectratio]{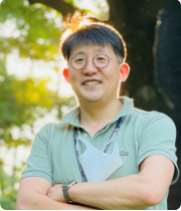}}]{Chansoo Kim}
    is an assistant professor at the University of Science and Technology and a senior research scientist at the Computational Science Centre, Korea Institute of Science and Technology. He leads the AI/R (AI, Information \& Reasoning) Lab., which explores the theoretical foundations of AI, the science of information, and complex (adaptive) systems. His researches span ethics and alignment, optimization, decentralization, and causality in AI—ranging from mathematical theory to real-world applications. Prof. Kim’s work centers on non-Gaussian behaviors—particularly heavy-tailed and leptokurtic—and their applications in learning, inference, finance, and inequality. He traces his academic lineage to C. F. Gau{\ss}. While grounded in theoretical AI, his lab’s research has also informed public policy. During the COVID-19 pandemic, the group supported the Korean CDC and the Office of the President with AI-driven, large-scale agent-based modeling.
\end{IEEEbiography}

\begin{IEEEbiography}[{\includegraphics[width=1in,height=1.25in,clip,keepaspectratio]{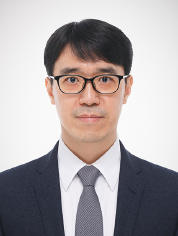}}]{Yun Jang}
    is a Professor of Computer Engineering at Sejong University, Seoul, South Korea. He received his Ph.D. in Electrical and Computer Engineering from Purdue University in 2007, specializing in data visual analytics, scientific visualization, and computer graphics. He also holds an M.S. from Purdue University (2002) and a B.S. in Electrical Engineering from Seoul National University (2000). Before joining Sejong University in 2012, Dr. Jang worked as a researcher at ETH Z\"{u}rich and the Swiss National Supercomputing Center. His research focuses on data visualization, visual analytics, artificial intelligence, and machine learning. He develops AI-driven solutions for volume rendering, social media analytics, smart city applications, traffic optimization, and healthcare monitoring. His work combines theoretical research with practical applications to help extract meaningful insights from complex data.
\end{IEEEbiography}

\begin{IEEEbiography}[{\includegraphics[width=1in,height=1.25in,clip,keepaspectratio]{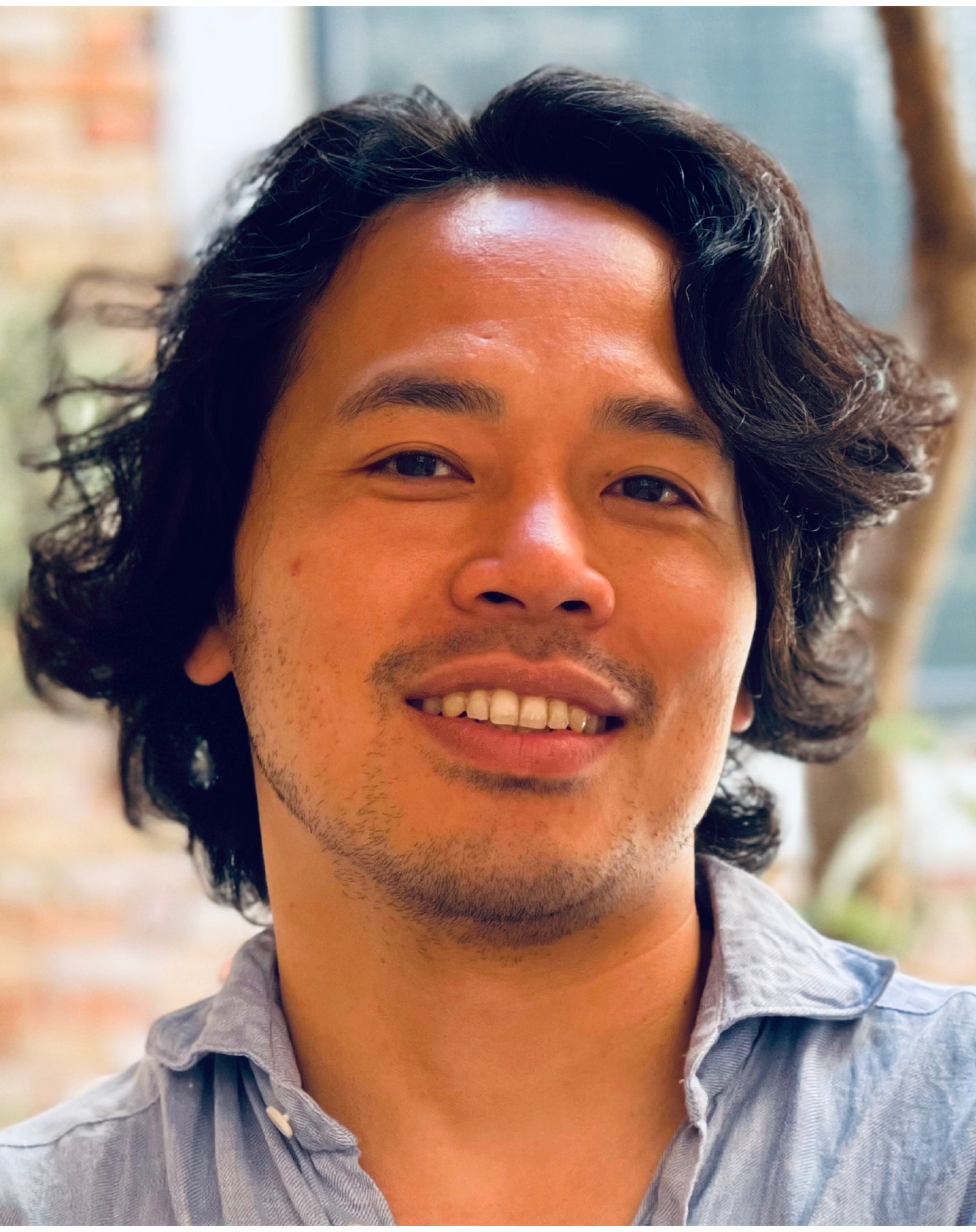}}]{Takanori Fujiwara} is an assistant professor at the Department of Computer Science at the University of Arizona, USA. His expertise spans visual analytics, machine learning, and network science, and he specializes in developing interactive dimensionality reduction techniques. He received his Ph.D. degree in Computer Science from UC Davis and his Master's and B.E. from the University of Tokyo. He was a postdoctoral researcher and a principal research engineer at the Department of Science and Technology at Linköping University. Prior to his Ph.D., he worked for Kajima Corporation in Japan.
\end{IEEEbiography}

\end{document}